\documentclass{aa}

\usepackage[varg]{txfonts}
\usepackage{graphicx}
\usepackage{xcolor}
\usepackage{amsmath}
\usepackage{natbib,twoopt}
\usepackage[breaklinks=true]{hyperref} 

\bibpunct{(}{)}{;}{a}{}{,} 
\makeatletter
\newcommandtwoopt{\citeads}[3][][]{\href{http://adsabs.harvard.edu/abs/#3}
{\def\hyper@linkstart##1##2{}
\let\hyper@linkend\@empty\citealp[#1][#2]{#3}}}
\newcommandtwoopt{\citepads}[3][][]{\href{http://adsabs.harvard.edu/abs/#3}
{\def\hyper@linkstart##1##2{}
\let\hyper@linkend\@empty\citep[#1][#2]{#3}}}
\newcommandtwoopt{\citetads}[3][][]{\href{http://adsabs.harvard.edu/abs/#3}
{\def\hyper@linkstart##1##2{}
\let\hyper@linkend\@empty\citet[#1][#2]{#3}}}
\newcommandtwoopt{\citeyearads}[3][][]
{\href{http://adsabs.harvard.edu/abs/#3}
{\def\hyper@linkstart##1##2{}
\let\hyper@linkend\@empty\citeyear[#1][#2]{#3}}}
\makeatother

\makeatletter
\let\linenumbers\relax

\makeatother

\begin{document}

\title{The contribution of neutral gas to Faraday tomographic data at low frequencies}

\subtitle{A first extensive comparison between real and synthetic data}

\author{J. Berat \inst{1,3}
    \and    M-A. Miville-Deschênes \inst{1,3}
    \and    A. Bracco \inst{2,1}
    \and    P. Hennebelle \inst{3}
    \and    J. Scholtys \inst{3}}

\institute{Laboratoire de Physique de l'ENS, Université Paris Cité, Ecole normale supérieure, Université PSL, Sorbonne Université, Observatoire de Paris, CNRS, 75005 Paris, France
\and INAF – Osservatorio Astrofisico di Arcetri, Largo E. Fermi 5, 50125 Firenze, Italy
\and AIM, CEA, CNRS, Université Paris-Saclay, Université Paris Cité, 91191, Gif-sur-Yvette, France}

\date{}

\abstract{LOFAR observations of diffuse interstellar polarization at meter wavelengths reveal intricate polarized intensity structures with an unexpected correlation with neutral HI filaments that could not be reproduced in simulations with low cold neutral medium (CNM) abundance.}
{We investigate whether magneto-hydrodynamic simulations of thermally bi-stable neutral interstellar medium, with a range of CNM fraction, can reproduce the properties of the 3C196 field, the high Galactic latitude test observational field.}
{Using 50 pc simulations with varying levels of turbulence and compressibility, we generated synthetic 21 cm and synchrotron observations, including instrumental noise and beam effects, for different line-of-sight orientations relative to the magnetic field. To this end we developed \texttt{MOOSE}, a code to generate synthetic synchrotron polarization and Faraday tomography. We also developed a metric, $\eta$, based on the HOG algorithm, to quantify the relative contribution of cold and warm neutral medium structures to the Faraday tomographic data.}
{The synthetic observations show levels of polarization intensity and $RM$ values comparable to the 3C196 field, indicating that thermal electrons associated with the neutral \ion{H}{I} phase can account for a significant fraction of the synchrotron polarized emission at 100–200 MHz.
The simulations consistently reveal a correlation between CNM and Faraday tomographic structures that depends on turbulence level, magnetic field orientation, and observational noise, but only weakly on CNM fraction. We found slightly weaker CNM-Synchrotron polarized emission correlation level than observed in the 3C196 field. Many elements could contribute to this result: $n_e$ prescription, contribution of the Warm Ionized Medium (not modeled), the longer line-of-sight in the data compared to the simulations, the Fourier driving producing CNM structures less aligned with $\vec{B}$. }
{These results, although based on comparison with the 3C196 field alone, suggest that low-frequency polarimetric observations provide a valuable probe of magnetic-field morphology in the multi-phase Solar-neighborhood ISM, while simultaneously underscoring the need for improved modeling of the turbulent, multi-phase, and partially ionized interstellar medium. A broader comparative analysis will be essential to more fully clarify how our findings inform the connection between low-frequency polarization data and the structure of partially ionized gas.}

\maketitle

\section{Introduction}
The transition from the warm neutral medium (WNM) to the cold neutral medium (CNM) is a critical step in the interstellar medium (ISM) cycle, pivotal for the formation of molecular clouds and stars. This transition, largely governed by the thermal instability \citep[e.g.,][]{Field1965, Wolfire2003, AuditHennebelle2005, Saury2014, Godard2024}, marks a key stage in the condensation of diffuse gas into denser, star-forming regions. Despite extensive theoretical and numerical investigations, the precise mechanisms regulating this phase change remain incompletely understood. Among the factors influencing this transition, magnetic fields have emerged as a possible player.

The polarization of thermal dust emission, extensively mapped by the \textit{Planck} satellite \citep{PlanckXXXII}, has offered high-resolution insights into the magnetic field morphology, particularly in relation to filamentary structures in the diffuse ISM. These studies revealed statistical alignments between magnetic fields and \ion{H}{I} filaments, suggesting a strong coupling between magnetic field orientations and the physical processes governing the ISM. Magnetic fields in the ISM have been increasingly recognized as a key factor in shaping its structure and dynamics, particularly through their interaction with neutral gas phases. Recent work by \citet{Clark2015,Clark2019} has used \ion{H}{I} observations and \textit{Planck} dust polarization data to reveal alignment between the morphology of \ion{H}{I} filaments and the orientation of magnetic fields. These studies provided evidence that filamentary structures in the CNM may trace the structure of the local magnetic field.

While starlight and dust polarization studies have advanced our understanding of the magnetic field's large-scale morphology, another proxy of the magnetic field has been less studied for local ISM. Synchrotron emission, arising from relativistic electrons spiraling along magnetic field lines, provides a complementary perspective. Synchrotron radiation traces the component of the magnetic field perpendicular to the line of sight (LOS), offering insights into its strength and orientation \citep{Rybicki1979, Longair2011,Padovani2021}. Observations at low radio frequencies are particularly valuable, as they capture Faraday rotation effects, wherein the plane of polarized light rotates as it propagates through a magnetized, ionized medium \citep{Burn1966, sokoloff98, Ferriere2021}. This effect depends on the thermal electron density and the magnetic field component along the LOS, making it a powerful tool for probing the magneto-ionic medium.
The degree of rotation experienced by a polarized wave is quantified by the Faraday depth, which depends on the thermal electron density, $n_e$, the magnetic field component along the LOS, $B_{\parallel}$, and $s$, the path length.

The technique of Faraday tomography \citep{Brentjens2005} exploits these low-frequency radio observations to decompose polarized synchrotron emission by Faraday depth, enabling a non-spatial 3D reconstruction of the magneto-ionic structure of the ISM\citep{Haverkorn2004, VanEck2018, Erceg2022, Erceg2024}. Recent advancements in radio facilities, such as the Low Frequency-Array \citep[LOFAR; e.g.][]{VanHaarlem2013} have significantly improved the sensitivity and resolution of low-frequency synchrotron polarization observations, opening new questions for the interplay between magnetic fields and the thermal instability.
 
A growing body of research seeks to connect these radio observables with neutral gas structures \citep[e.g.,][]{Vaneck2017}. Observationally, \citet{Bracco2020} demonstrated that CNM structures, extracted via phase decomposition of 21~cm data with the Regularized Optimization for Hyper-Spectral Analysis (\texttt{ROHSA}) code \citep{Marchal2019}, exhibit stronger spatial correlations with Faraday tomography data compared to WNM structures. However, numerical simulations by \citet{Bracco2022} reported an inverse trend, with WNM showing greater correlation with synchrotron data. This discrepancy was attributed to the low CNM mass fraction ($1~\%$) in the simulated environment and the absence of instrumental effects in the analysis, highlighting the need for a more systematic investigation. Besides, depolarization canals, which are narrow regions of reduced polarized intensity observed in LOFAR data \citep[e.g.,][]{Jelic2015, Erceg2022}, have been shown to align with filamentary \ion{H}{I} structures in the CNM \citep{Jelic2018, Clark2019}, further supporting a close link between small-scale magnetized structures in the neutral ISM and features seen in polarization observations \citep[e.g.,][]{Haverkorn2004, Jelic2015}. In particular, this behavior was observed in the 3C196 field, which is used in this study to compare with. However, it represents only a single LOS through the Galaxy and therefore cannot fully capture the morphological and physical diversity of the diffuse ISM.

The study presented here revisits these questions by using a suite of magneto-hydrodynamic (MHD) simulations \citep{Rashidi2024} that spans a range of CNM mass fractions (10~\%–50~\%), providing a more representative depiction of the diffuse ISM. In addition, we incorporate a simplified instrument model to simulate telescope beam effects and instrumental noise, ensuring a more realistic comparison between simulations and observations. 

To achieve this, we developed a Julia-based \citep{Julia2017} code, Mock Observation Of Synchrotron Emission (\texttt{MOOSE}), to produce synthetic observations of polarized synchrotron emission from MHD simulations, perform Faraday tomography, and compute spatial correlations between \ion{H}{I} phases and Faraday observables. This approach seeks to test whether low-frequency synchrotron emission can provide new insights into magnetic field-driven effects in ISM phase transitions, bridging the gap between theoretical predictions and observational constraints.

The paper is organized as follows: Sect.~\ref{sec: 3C196 field of view} is a review of the physical properties of the 3C196 field, which has been used throughout the analysis for comparison with numerical simulations. In Sect.~\ref{sec: methods}, we present the methods used to analyze the correlation between Faraday tomography observables and neutral gas structures. This includes the Histogram of Oriented Gradients (HOG) method \citep{Soler19}, the process of generating synthetic synchrotron polarization maps using the \texttt{MOOSE} code, and the impact of instrumental effects such as noise filtering and beam convolution. In Sect.~\ref{sec: Results}, we present our findings, focusing on the dependence of the \ion{H}{I}-Faraday correlation on CNM fraction, turbulence, and magnetic field topology. We compare the results of different simulation setups and contrast them with the LOFAR Two-metre Sky Survey (LoTSS) observations of the 3C196 field \citep{Erceg2022}. In Sect.~\ref{sec: discussion}, we examine the limitations of our simulations and explore possible physical processes, such as electron density prescription, cosmic ray transport and supernova-driven turbulence, that may influence Faraday tomography observables and their correlation with \ion{H}{I} data. Finally, in Sect.~\ref{sec: conclusions}, we summarize the key findings.

\section{The 3C196 field of view}
\label{sec: 3C196 field of view}

\begin{figure}
\centering
\includegraphics[width=0.95\linewidth]{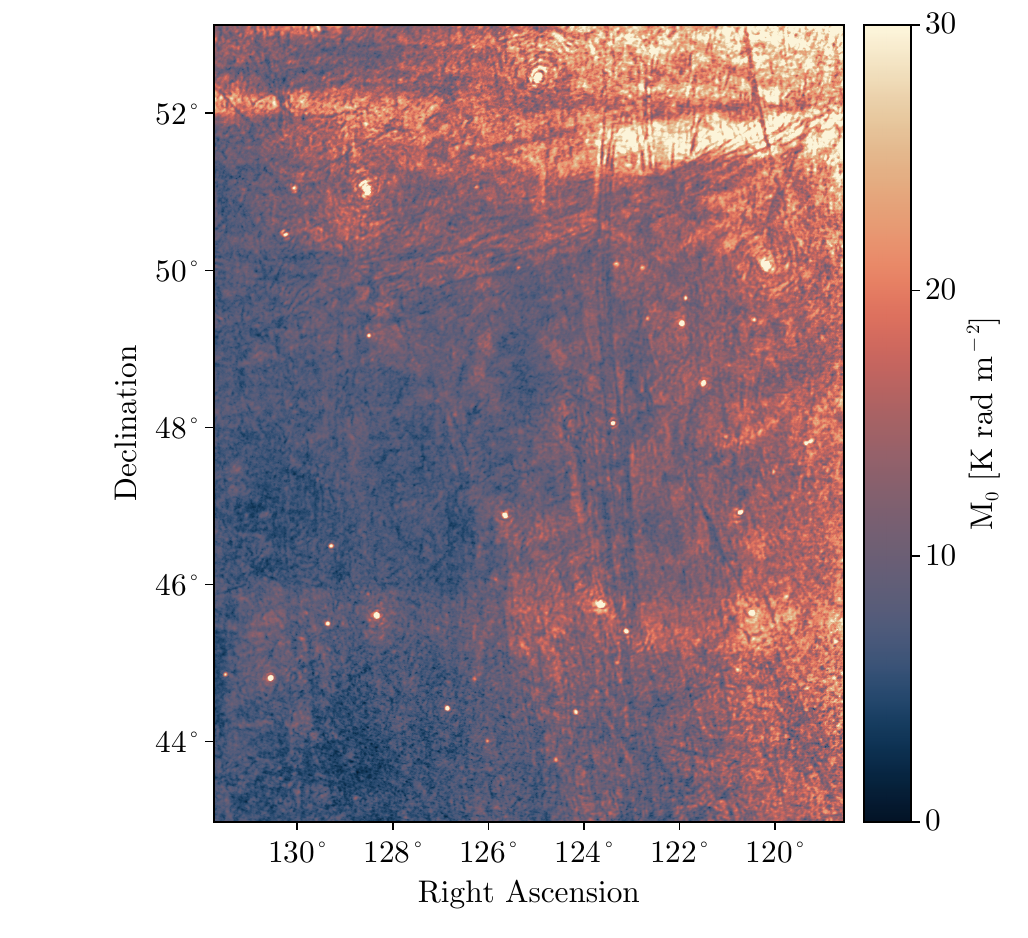}
\includegraphics[width=0.95\linewidth]{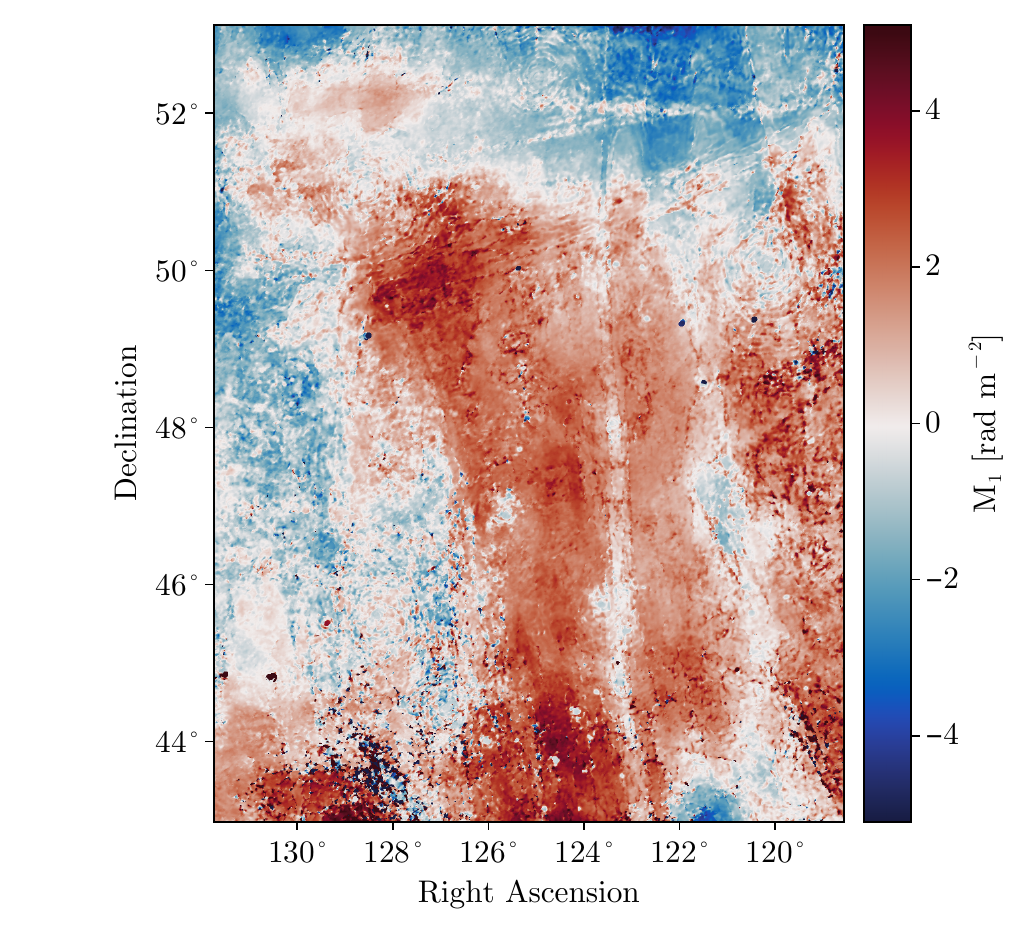}
\includegraphics[width=0.95\linewidth]{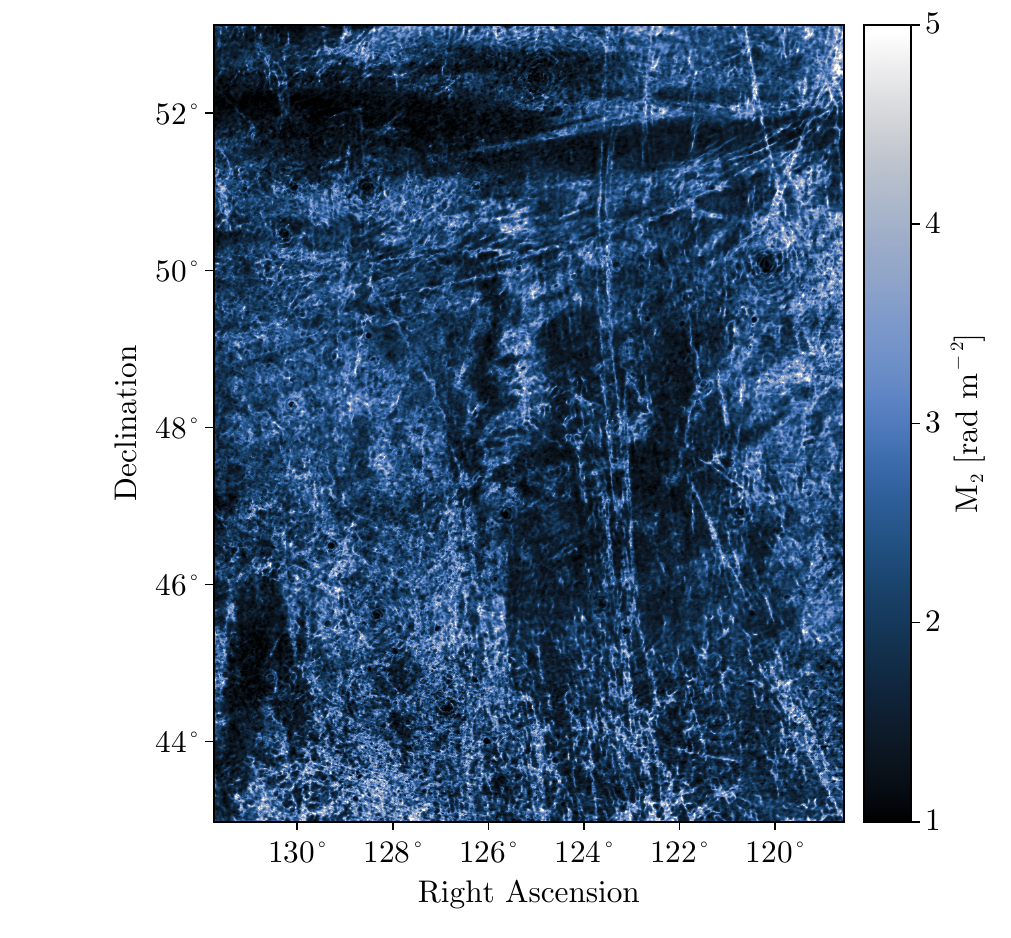}
\caption{Moment 0 (integrated polarized intensity) de-biased for noise (\textit{top}), Moment 1 (\textit{middle}) and Moment 2 (\textit{bottom}) of the 3C196 field from LOFAR LoTSS DR2 Faraday tomographic data \citep{Erceg2022}.}
\label{fig: moments_3C196}
\end{figure}

\subsection{Observations}
The 3C196 field is one of the primary fields of view of the LOFAR-Epoch of Reionization key science project. It is located in a diffuse region at high Galactic latitude ($l = 171^\circ$ , $b =+33^\circ$) centered on the extragalactic source 3C196 \citep{Jelic2015,Jelic2018}. The angular size of the field is $64~\textrm{deg}^2$. Observations were done between 115 and 189 MHz with 3.2 kHz spectral resolution. It is at the core of the analysis of \citet{Bracco2020} that revealed morphological correlations between low frequencies radio polarization and \ion{H}{I} data from Effelsberg-Bonn \ion{H}{I} survey \citep[EBHIS;][]{Kerp2011,winkel2016}. In their study \citet{Bracco2020} used an early single pointing version of the 3C196 field LOFAR observations which shows a sharp increase in noise at the edges of the field of view. Later on \citet{Erceg2022} presented a complete Faraday tomographic mosaic of the LoTSS survey conducted with the LOFAR telescope, that includes the 3C196 field. Observations were conducted between 120 and 168 MHz with a resolution of 97.6 kHz. The mosaicking strategy, involving overlapping fields of view, results in a more uniform noise level across the observed area. 

In this work  we reproduce the data analysis methodology of \citet{Bracco2020}. To do so we use the LOFAR data product of \citet{Erceg2022} and, like \citet{Bracco2020}, the 21\,cm data from EBHIS. 
The data products have a respective angular resolution of $4'$ for LOFAR and $10'8$ for EBHIS.
To compare the two datasets, the LOFAR polarimetric data cubes have been convolved at the EBHIS resolution, then projected onto the EBHIS grid.

\subsection{Physical properties of the 3C196 field}

In this section, we describe the general physical properties of the 3C196 field, with which we will compare to the simulations. We are mostly interested in the multi-scale properties of the \ion{H}{I} data and the intensity of the magnetic field.

\begin{figure*}[!ht]
\centering
\includegraphics[width=\linewidth]{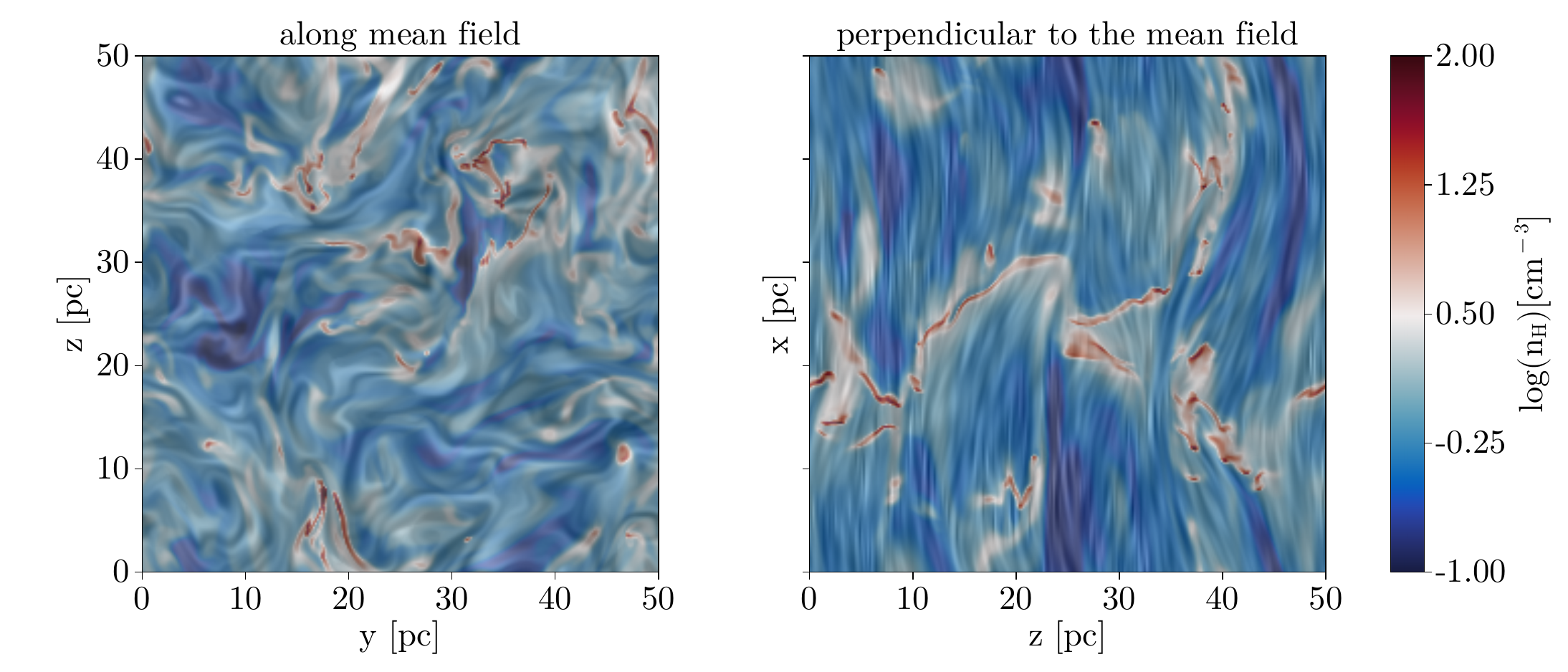}
    \caption{Slice through the simulation cube showing the logarithmic gas density overlaid with magnetic field orientation using the line integral convolution (LIC) technique. \textit{Left:} projection along the mean magnetic field direction; \textit{Right:} projection perpendicular to the mean field.}
    \label{fig: LIC along and perp}
\end{figure*}

Based on EBHIS 21\,cm data \citep{winkel2016}, the average \ion{H}{I} column density in the 3C196 field is $N_{\ion{H}{I}}=4.2 \times 10^{20}\,$cm$^{-2}$. 
Using on the HI phase decomposition done in \citet{Bracco2020} using \texttt{ROHSA} \citep{Marchal2019}, the WNM column density within the field of view is $N_\textrm{WNM}$, to be $2.4 \times 10^{20}~\textrm{cm}^{-2}$. The fraction of the \ion{H}{I} in the CNM phase, defined as $f_{\rm CNM} = N_{\rm CNM}/N_{\ion{H}{I}}$, is $38\%$. Moments 0, 1 and 2 of the Faraday spectra of the 3C196 field are displayed in Fig.~\ref{fig: moments_3C196}.

The \texttt{ROHSA} decomposition also allows estimating the \ion{H}{I} velocity dispersion along the LOS. Here we take the WNM as the reference as it is volume filling. The WNM velocity dispersion $\sigma_\textrm{u}$ in the 3C196 field is $16.3~\textrm{km~s}^{-1}$.
This velocity dispersion results from the combination of thermal and turbulent contributions, and can be expressed as:  
$\sigma^2_u = \sigma_\textrm{turb}^2 + \sigma_\textrm{therm}^2.$
Assuming a reference temperature of 7000~K for the WNM, the thermal velocity dispersion is  $\sigma_\textrm{therm} \approx 7.6~\textrm{km s}^{-1}$ leading to a turbulent velocity dispersion of $\sigma_\textrm{turb} \approx 14.4~\textrm{km~s}^{-1}.$
Following \citet{Kolmogorov41}, the turbulent velocity dispersion related to scale $l$ like:
\begin{equation}
\label{eq:K41}
     \sigma_{\textrm{turb}} = \sigma_{u,1\textrm{pc}}\left(\frac{l}{1~\textrm{pc}}\right)^{1/3},
\end{equation}
where $\sigma_\textrm{u,1~\textrm{pc}}$ is the 1D turbulent velocity dispersion along the LOS at a scale of 1 pc.

According to the 3D-dust map of \citet{Edenhofer2024}, the Galactic interstellar matter observed in the 3C196 field is distributed over a depth of $\sim 350$~pc, located approximately 200~pc from the Sun, corresponding to the wall of the Local Bubble.
Assuming the WNM is filling the whole depth of 350~pc, we get $\sigma_\textrm{u,1~pc} = 2.0~\textrm{km~s}^{-1}$. 

At the 50~pc scale (the simulation scale), the extrapolated 1D velocity dispersion is $\sigma_\textrm{u,50~pc} = 7.4~\textrm{kms}^{-1}$, yielding a 3D turbulent velocity dispersion of $\sigma_{3\textrm{D}} = 12.8~\textrm{km~s}^{-1}$.

According to \citet{Heiles2005}, the magnetic field strength at high Galactic latitude is estimated at $6 \pm 2~\mu\textrm{G}$ based on 21~cm Zeeman splitting measurements. Previous analyses of this region \citep{Jelic2015, Turic2021, Erceg2022} indicate that the magnetic field is largely dominated by its plane-of-the-sky component, with only a modest contribution along the LOS. Therefore, the observed Faraday depths do not necessarily imply locally enhanced LOS magnetic-field strengths, but rather result from the interplay of electron density and path length along this particular LOS.

\begin{table}[ht!]
\caption{Description of the simulations used in this study.}
\label{table:1}
\centering
\begin{tabular}{c c c | c c c c c} 
\hline\hline
N & $\chi$ & $\textrm{A}_\textrm{f,RMS}$ & $B_\mathrm{RMS}$ & $u_\mathrm{RMS}$ & f$_{\textrm{CNM}}$ & f$_{\textrm{vCNM}}$ & $\sigma_{u,1\textrm{pc}}$ \\
 & & & $\mu\mathrm{G}$ & $\mathrm{km/s}$& \% & \% & $\mathrm{km/s}$ \\
\hline
1  & 0.0 & 18000 & 2.1 & 7.7  & 17 & 1.9 & 1.2 \\
2  & 0.2 & 9000 & 1.0 & 5.2  & 23 & 2.3 & 0.8 \\
3  & 0.2 & 18000 & 1.8 & 7.4  & 19 & 2.0 & 1.2 \\
4  & 0.2 & 36000 & 3.4 & 9.9  & 18 & 1.9 & 1.6 \\
5  & 0.2 & 72000 & 5.8 & 14.5 & 20 & 3.0 & 2.3 \\
6  & 0.5 & 9000 & 0.9 & 4.8  & 31 & 3.6 & 0.8 \\
7  & 0.5 & 18000 & 1.9 & 6.9  & 29 & 3.3 & 1.1 \\
8  & 0.5 & 36000 & 3.1 & 9.9  & 26 & 2.7 & 1.6 \\
9  & 0.5 & 72000 & 4.6 & 13.0 & 26 & 4.2 & 2.0 \\
10 & 0.8 & 9000 & 0.8 & 4.3  & 39 & 4.4 & 0.7 \\
11 & 0.8 & 18000 & 1.5 & 6.1  & 35 & 4.1 & 0.9 \\
12 & 0.8 & 36000 & 2.5 & 8.3  & 42 & 5.3 & 1.3 \\
13 & 0.8 & 72000 & 4.2 & 12.0 & 40 & 6.3 & 1.8 \\
14 & 1.0 & 18000 & 1.7 & 6.8  & 46 & 5.1 & 1.0 \\
\hline
\end{tabular}
\tablebib{For all simulations, the box is 50\,pc aside, the mean density is $\langle n_H \rangle = 1\,$cm$^{-3}$, corresponding to an average column density of $\langle N_{\textrm{HI}} \rangle = 1.55 \times 10^{20}\,$cm$^{-2}$. The mean $\vec{B}$-field intensity is $7.64~\mathrm{\mu G}$.
Input parameters of the simulations: $\chi$, the ratio between the compressible and solenoidal turbulent modes; $A_\textrm{f,RMS}$, the initial forcing amplitude. Measured quantities after steady state was reached: 
$B_\mathrm{RMS}$, the RMS value of the $\vec{B}$-field; $\vec{u}_\mathrm{RMS}$, the RMS value of the velocity field; $\textrm{f}_\textrm{CNM}$, the mass fraction of CNM; $\textrm{f}_\textrm{vol,CNM}$ the volume fraction of CNM and $\sigma_{u,1\textrm{pc}}$, the velocity dispersion at 1 pc.  }
\end{table}

\section{Methods}
\label{sec: methods}

\subsection{Description of numerical simulations}
\label{sec: Description_NumSim}
In this study, we use ideal-MHD simulations designed with the \texttt{RAMSES} code \citep{teyssier2022, fromang2006} that uses adaptive mesh refinement (AMR) up to $1024^3$. We use a version of these simulations computed on a cartesian grid of $256^3$ pixels. The box physical size is $L = 50$ pc aside, leading to a resolution of $\Delta x \approx 0.20$ pc. 

The physical conditions we want to reproduce are the ones of atomic gas clouds or those transitioning to molecular gas like what is expected on the walls of the Local Bubble \citep{Inoue&Inoutsuka16}.
In this diffuse medium ($0.1$ to $100~\textrm{cm}^{-3}$, \citet{Ferriere2001}), the self-gravity energy density is significantly lower (two to three orders of magnitude) than turbulent and magnetic energy densities, hence self-gravity is excluded from the simulations. The cooling function is based on \citet{AuditHennebelle2005} and closely follows \citet{Wolfire2003}. It includes photo-electric heating of dust grains and cooling by electron recombination onto dust grains, as well as collisional excitation of \ion{Ly}{$\alpha$}, \ion{C}{II}, and \ion{O}{I} emissions.

To maintain turbulence in the simulation, a Fourier space forcing is modeled by an Ornstein-Ulhenbeck stochastic process \citep{Eswaran1988, Saury2014}. A turbulent velocity field is added at large-scale ($1 < k < 3$), peaking at $k = 2$ (half the box size L). The forcing is parametrized by its amplitude, $A_\textrm{f,RMS}$, and the compressive to solenoidal modes decomposition parameter, $\chi$, where $\chi = 0$ only contains solenoidal modes and $\chi = 1$ only compressible modes. 

In this study, we use a set of 14 simulations described in Table~\ref{table:1}. The mean $\vec{B}$-field intensity of all simulations is $7.64~\mathrm{\mu G}$. The ratio of compressible to solenoidal modes of the forcing was varied from $\chi =0 $ to $\chi =1$, and its amplitude, $A_\textrm{f,RMS}$, was varied between 9000 and 72000 producing 3D velocity dispersions from 4.3 and 14.5 km\,s$^{-1}$.

Each simulation begins with WNM conditions with a density of $n_H = 1~\mathrm{cm}^{-3}$ and a temperature of $T = 8000~\mathrm{K}$. The magnetic field is initiated uniformly along the $x$-axis, which gives a prevailing direction for the $\vec{B}$ field in the simulation box; $\nabla\cdot\vec{B} = 0$ is guaranteed by the use of the constrained transport method. Due to the nature of the periodic box, the coherence of the magnetic field along the $x$-axis is maintained at each timestep.

As the simulations evolve, shear flows and shock waves emerge in the volume, causing local pressure variations in the WNM leading to the formation of CNM structures. We let simulations proceed until a steady state is achieved when velocity dispersion, gas-phase mass fractions, density, and temperature PDFs reach a stationary state, within at least three dynamical timescales, $t_{\textrm{dyn}} = L/\sigma_{u,3D}$, where $\sigma_{u,3D}$ is the velocity dispersion in 3D. 
An illustration of one simulation is given in Fig.~\ref{fig: LIC along and perp}. where two slices of the density field are shown, one along and one perpendicular to the mean magnetic field, represented here using the Line Integral Convolution (LIC) technique \citep{Cabral1993}.

\begin{figure}
\centering
   \includegraphics[width=\linewidth]{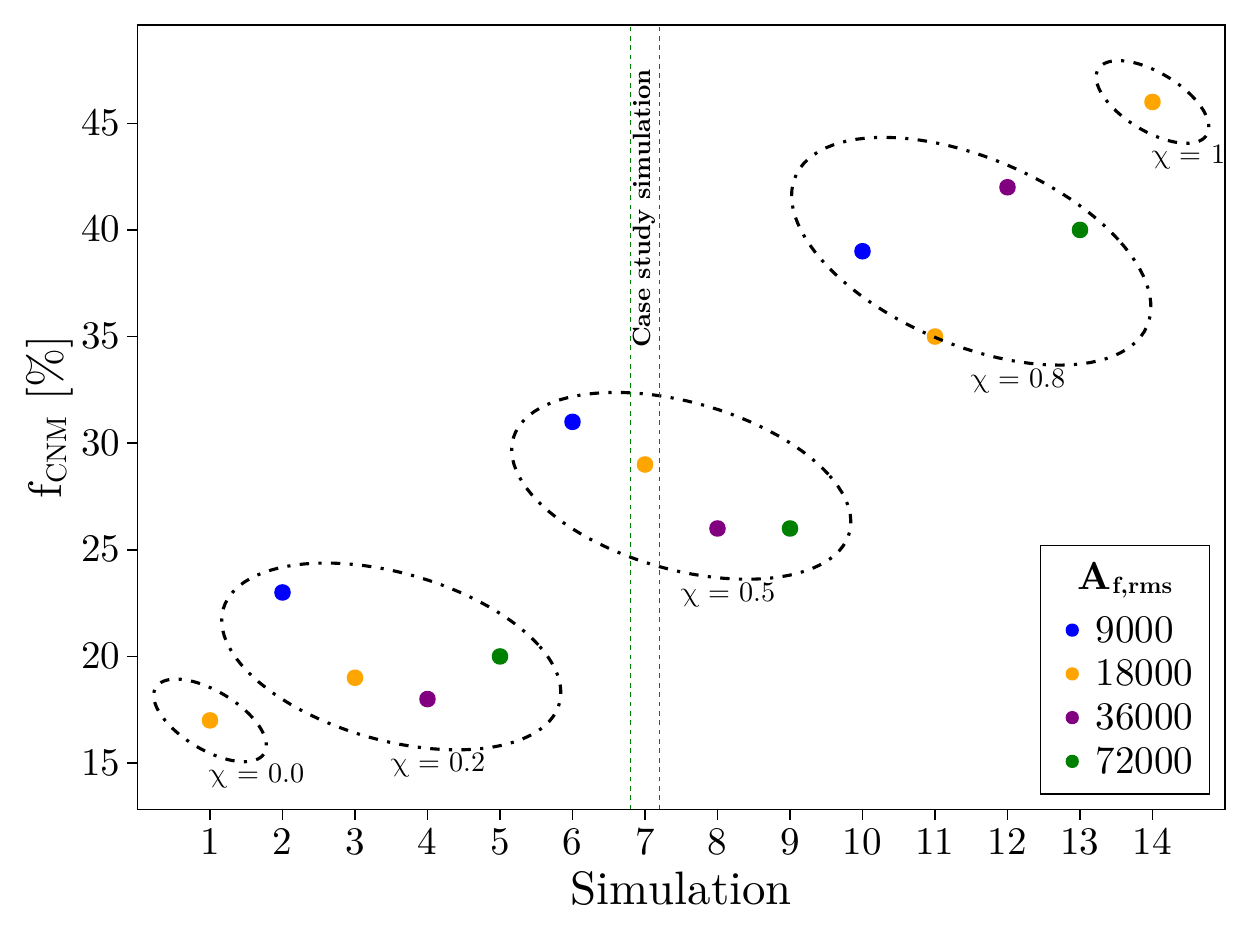}
     \caption{CNM mass fraction $\mathrm{f_{\textrm{CNM}}}$ of the different simulations used in this study. The dashed circle encircle the simulations with the same turbulent modes ratio $\mathrm{\chi}$. Colors show the initial input for the forcing amplitude.}
     \label{fig: CNM mass fraction plot}
\end{figure}

For each simulation, we decomposed the gas in the three main \ion{H}{I} components: CNM, Lukewarm Neutral Medium, LNM, and WNM. Like in \citet{Saury2014}, we notice that CNM structures form preferentially in over-pressured and less turbulent WNM regions. The overall CNM mass fraction in our set of simulations tends to increases with the rate of compressible modes and more so for less turbulent ones. This is illustrated in Fig.~\ref{fig: CNM mass fraction plot}, where the CNM mass fraction is plotted for the last timestep of each simulation. 
Table~\ref{table:1} also provides the CNM mass and volume fractions for the last timestep of each simulation, together with the root-mean-squared (RMS) values for $B_\textrm{RMS}$, $u_\textrm{RMS}$ and the velocity dispersion $\sigma_{u,1\textrm{pc}}$ of the gas at $1~\textrm{pc}$. 

\begin{figure}
\centering
   \includegraphics[width=\linewidth]{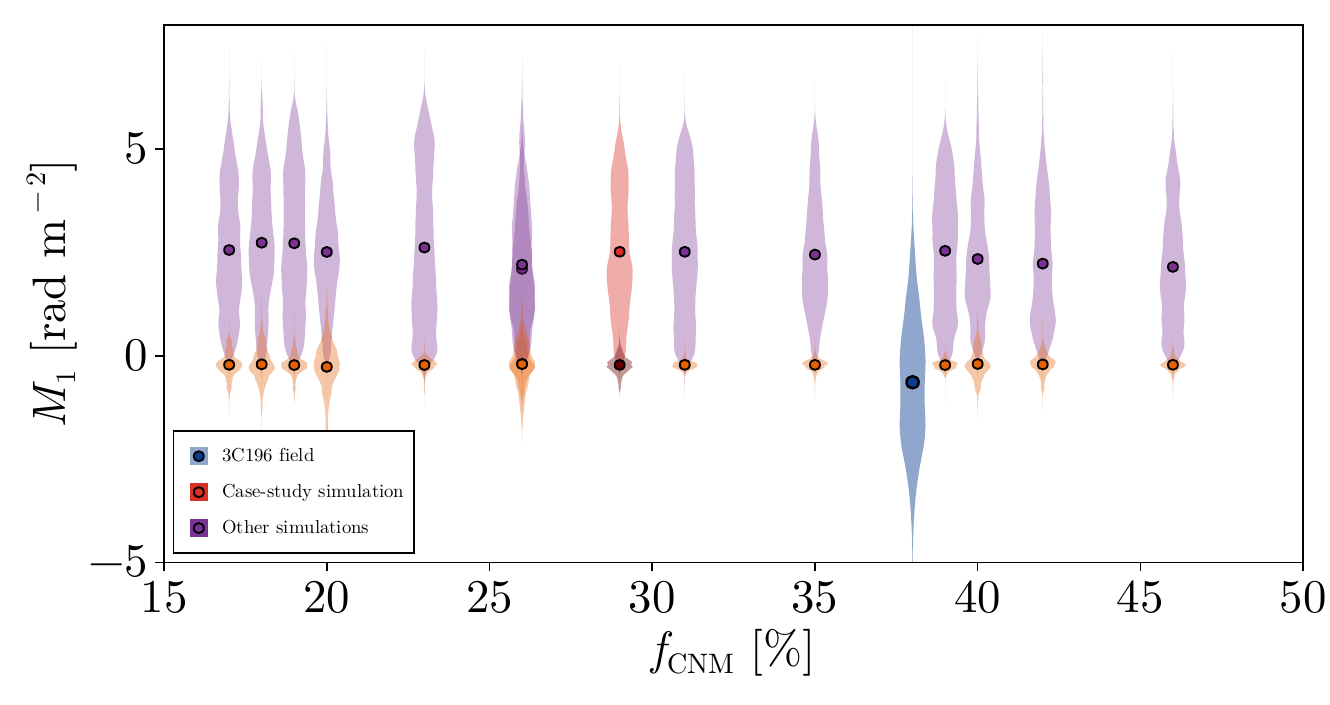}
     \caption{$M_1$ distributions as a function of $f_{\mathrm{CNM}}$ for the set of simulations. These distributions are measured along two orthogonal directions: parallel (lighter color) and perpendicular (darker color) to the mean magnetic field. Filled circles indicate the corresponding mean values $\langle M_1 \rangle$, outlined in black. The case-study simulation is highlighted in red, while the remaining simulations are shown in violet. Two simulations have $f_\textrm{CNM} = 26 \%$, leading to an over-plotting of their respective distributions. The 3C196 field is displayed in blue, with its associated distribution and mean value.}
     \label{fig: M1 all}
\end{figure}

In this paper we often refer to simulation 7 that we used as our case-study throughout our analysis. This simulation has a CNM fraction of $29\%$ and $\sigma_{u,\text{1pc}}=1.1\,\mathrm{km\,s^{-1}}$ both slightly lower than the values of the 3C196 field but still representative of the diffuse ISM. Moreover, the Faraday depth distributions of the simulations and of the 3C196 field, shown in Fig.~\ref{fig: M1 all}, exhibit comparable order of magnitude, namely a few rad m$^{-2}$.

\subsection{Synthetic observations}
\label{sec: synth obs}

In order to reproduce the analysis of \cite{Bracco2020} that compared polarized synchrotron and 21 cm emission data of the 3C196 field, we need to create corresponding synthetic observations for each numerical simulation. This section describes how these were created, including the effects of Faraday rotation and LOFAR beam on the polarized synchrotron emission.

\subsubsection{Synchrotron emission}
\label{sec: synchrotron emission}
In order to produce synthetic observations of synchrotron emission and its polarization, which is at the basis of the Faraday tomography, we use Eqs.~$1$ to $9$ of \citet{Padovani2021} derived from \citet{Ginzburg1965}. 
As cosmic ray electrons (CRe) travel through the ISM, they lose energy due to interactions with matter, magnetic fields, and radiation \citep{Longair2011}. These interactions modify the cosmic ray energy spectrum, $j_e(E)$, where $E$ represents the CRe energy.

For the CRe energy spectrum, $j_e(E)$, we use the same power-law definition as in 
\citet{Padovani2018,Orlando2018} :
\begin{equation}
\label{eq: j_e}
    j_e(E) = j_0 \frac{E^a}{(E+E_0)^b},
\end{equation}
where $j_0 = 2.1\times10^{18}~\textrm{eV}^{-1}\textrm{s}^{-1}\textrm{cm}^{-2}\textrm{sr}^{-1}$, $E_0 = 710$ MeV, $a = -1.3$, and $b = 1.9$. In our case, $j_e(E)$ is considered spatially constant in the simulated box.

Noting $m_e$ the electron mass, $e$ the electron charge, $c$ the speed of light, the synchrotron specific emissivity at a position $s$ can be divided into two linearly polarized components: one parallel and one perpendicular to the magnetic field component orthogonal to the line-of-sight, $\Vec{B}_\perp$:
\begin{equation}
\label{eq: epspara}
    \varepsilon_{\nu}^{\parallel}(s) = \frac{\sqrt{3}e^3}{2m_ec^2}\int\limits_{m_ec^2}^{+\infty}\frac{j_e(E)}{v_e(E)}B_\perp (s)(F(x)-G(x))dE,
\end{equation}
and
\begin{equation}
\label{eq: epsperp}
    \varepsilon_{\nu}^{\perp}(s) = \frac{\sqrt{3}e^3}{2m_ec^2}\int\limits_{m_ec^2}^{+\infty}\frac{j_e(E)}{v_e(E)}B_\perp (s)(F(x)+G(x))dE,
\end{equation}
where $x = \frac{\nu}{\nu_c}$ and $\nu_c$ is defined as the frequency at which the synchrotron intensity of CRe at energy $E$ is the greatest:
\begin{equation}
\label{eq: critfreq}
\nu_c(B(\vec{r}),E) = \frac{3e}{4\pi m_ec}B_\perp(\vec{r})\left(\frac{E}{m_ec^2}\right).
\end{equation}

Synchrotron emission is then :
\begin{equation}
\label{eq: Inu}
    I_\nu(\vec{s}) = \int\limits_0^L \varepsilon_{\nu}^{\parallel}(\vec{s}) + \varepsilon_{\nu}^{\perp}(\vec{s}) d(\vec{s}).
\end{equation}
\subsubsection{Faraday rotation}
\label{sec: Faraday Rotation}

In this paper, we address the scenario where Faraday rotation is completely intertwined with synchrotron emission, leading to differential Faraday rotation \citep[e.g.][]{sokoloff98} as in \citet{Bracco2022}. Each layer within the simulated cubes contributes to both synchrotron emission and Faraday rotation. Stokes $Q_\nu$ and $U_\nu$ for synchrotron emission are given by the following equations :
\begin{equation}
\label{eq: Unu}
    U_\nu = \int\limits_0^L(\varepsilon_{\nu}^{\parallel}(s)-\varepsilon_{\nu}^{\perp}(s))\sin{\left(2(\Psi_{\textrm{loc}}(s) + \Delta\Psi_\nu(s))\right)}~ds,
\end{equation}
and
\begin{equation}
\label{eq: Qnu}
    Q_\nu = \int\limits_0^L(\varepsilon_{\nu}^{\parallel}(s)-\varepsilon_{\nu}^{\perp}(s))\cos{\left(2(\Psi_{\textrm{loc}}(s) + \Delta\Psi_\nu(s))\right)}~ds,
\end{equation}
where $\Psi_{\textrm{loc}}$ is the polarization angle at the position $s$ defined as the orientation of $\vec{B}_\perp$ rotated by $90^\circ$.
$\Delta \Psi_\nu$ describes the Faraday rotation, which is the integral along the LOS between the observer and $s$. The Faraday rotation depends on the squared of the observation wavelength $\lambda$, the electron density $n_e$ and the magnetic field along the LOS. It is also expressed as
\begin{equation}
\label{eq: Faraday angle}
    \Delta\Psi_\nu(s) = \phi(s)\lambda^2,
\end{equation}
where $\lambda = \frac{c}{\nu}$ and $\phi$ is the Faraday depth, a quantity independent of wavelength that describes the physical quantities responsible for the angle rotation from the emitting source towards the observer, corresponding to the cumulative Faraday rotation experienced by the polarized emission along the LOS :  
\begin{equation}
\label{eq: phi}
    \frac{\phi(s)}{[\mathrm{rad~m}^{-2}]} = 0.81\int\limits_0^{s} \frac{n_e}{[\mathrm{cm}^{-3}]} \frac{\vec{B}}{[\mu \mathrm{G}]}\cdot\frac{ds}{[\mathrm{pc}]},
\end{equation}
where $ds = \Delta x$ in the discrete case of our numerical simulations.
From Equations~\ref{eq: Unu} and \ref{eq: Qnu}, we define the polarized intensity $P$ as :
\begin{equation}
\label{eq: P}
    P = Q_{\nu} + i U_{\nu}.
\end{equation}
$P$ is commonly computed as an analogous of the Fourier transform into the Faraday spectrum:
\begin{equation}
\label{eq: Fphi}
    F(\phi) = \frac{1}{\pi}\int\limits_{\lambda_\mathrm{min}^2}^{\lambda_\mathrm{max}^2} P(\lambda^2)e^{-2i\phi\lambda^2}d(\lambda^2),
\end{equation}
For the limit case, where the integration is made all along the LOS, i.e. $s = 0$ and $s+ds = L$, $\phi$ becomes the rotation measure ($RM$).

\begin{figure}
\centering\includegraphics[width=\linewidth]{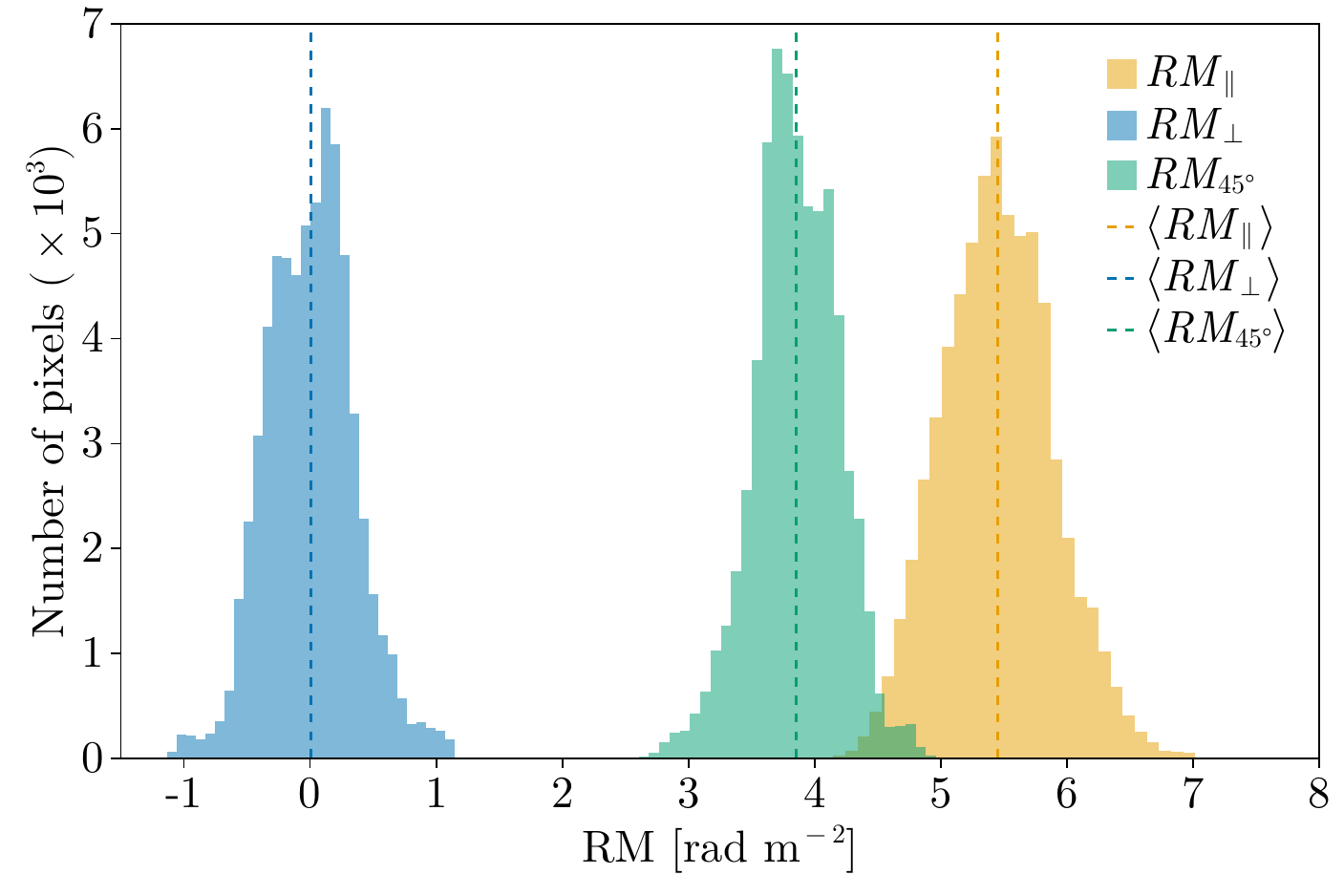}
     \caption{Simulation 7. $RM$ histograms for cubes produced with the line-of-sight parallel, perpendicular or at $45^\circ$ with the mean direction of $\vec{B}$. \textit{Dashed lines}: the mean values of the rotation measure maps. The mean values are $\langle RM_\parallel\rangle = 5.5~\textrm{rad~m}^{-2}$, $\langle RM_\perp\rangle = 0.0~\textrm{rad~m}^{-2}$, $\langle RM_{45^\circ}\rangle = 3.9~\textrm{rad~m}^{-2}$.}
\label{fig: RM_histogram}
\end{figure}

The Faraday spectrum is represented as polarized brightness temperature by taking the modulus of $F(\phi)$ making it a non-negative quantity. From now on, we will use $F(\phi)$ as representative of the modulus of Eq.~\ref{eq: Fphi}.

\subsubsection{Electron density}
In order to produce synthetic synchrotron observations including Faraday rotation, we need to estimate the density of thermal electrons, $n_e$, at each position in the cube. The numerical simulations are thus post-processed to estimate $n_e$ in units of $\mathrm{cm}^{-3}$
following Eq.6 of \citet{Bracco2022}, which derives from \citet{Wolfire2003} and \citet{Bellomi2020}:
\begin{equation}
\label{eq: ne}
    n_e \approx 2.4 \times 10^{-3}\zeta_{16}^{0.5}T_2^{0.25}\frac{G_{\mathrm{eff}}^{0.5}}{\phi_{\mathrm{PAH}}} + n_\textrm{H} X_{C^+},
\end{equation}
where $\zeta_{16}$ represents the total ionization rate per hydrogen atom due to energetic photons (including extreme UV and soft X-ray) and CRs expressed in units of $10^{-16}~\textrm{s}^{-1}$, $T_2$ is the temperature of the gas expressed in units of  $100~\textrm{K}$, $\phi_{PAH}$ denotes the recombination rate of electrons onto small dust grains such as polycyclic aromatic hydrocarbons (PAH). Furthermore, $X_{\mathrm{C}^+}$ is the relative abundance of ionized carbon compared to the number of hydrogen atoms ($n_H$). Following \citet{Bracco2022}, we use $X_{C^+} = 1.4~\times 10^{-4}$ and  $\phi_\mathrm{PAH} = 0.5$, $G_{\mathrm{eff}} = 1$, and $\zeta = 2.5~\times 10^{-16}~\mathrm{s}^{-1}$ \citep{Padovani2022} as representative values for the diffuse ISM. The impact of these choices on our results are discussed in Sect.~\ref{sec: discussion}. 

\begin{figure*}[!ht]
    \centering
    \includegraphics[width=\linewidth]{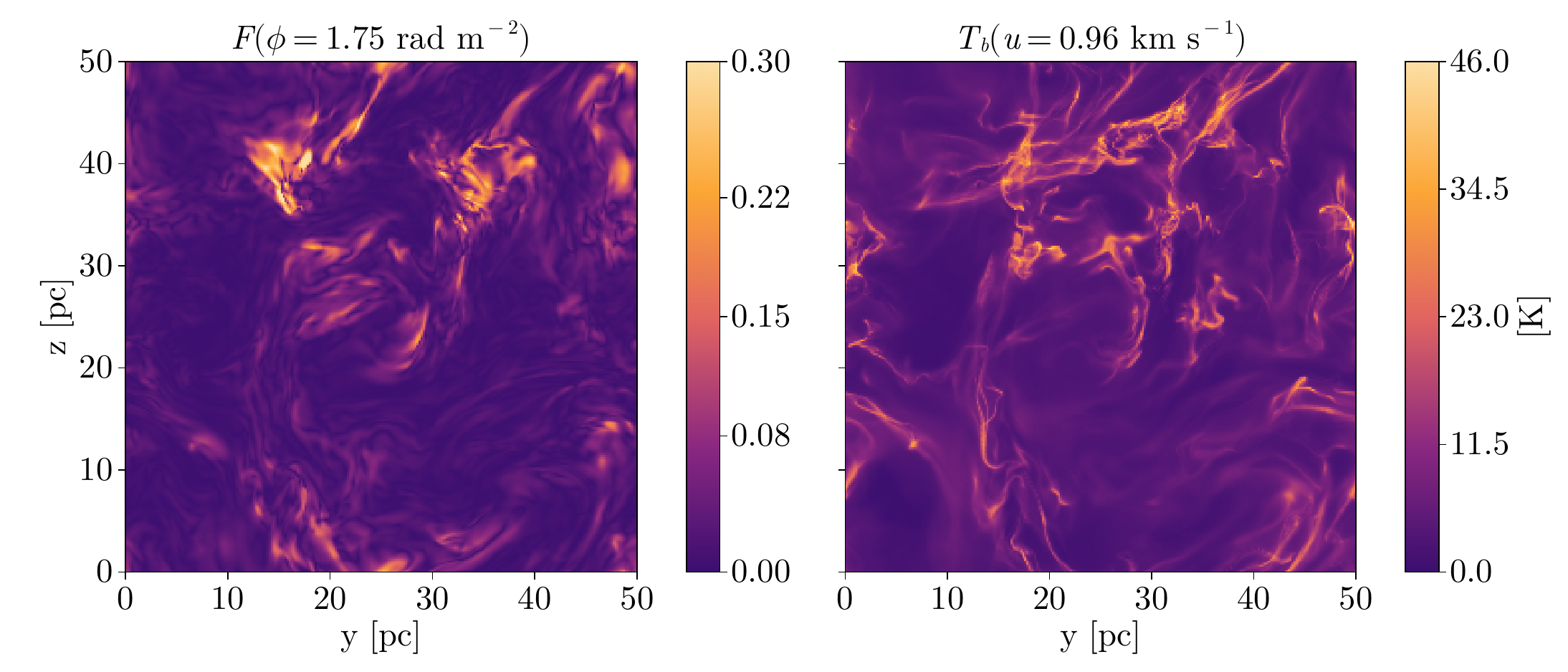}
    \caption{Mock observations for simulation 7 in the case where the line-of-sight is parallel to the mean $\vec{B}$ field. \textit{Left panel:} Map of the Faraday spectrum $F(\phi)$ at $\phi=1.75~\mathrm{rad~m^{-2}}$, \textit{Right panel:} Map of the 21\,cm brightness temperature $T_B(u)$ for $v=0.9~\mathrm{km~s^{-1}}$.}
    \label{fig: Fphi-Tb_instrument}
\end{figure*}

\subsubsection{Rotation Measure}

Before computing mock LOFAR observations, it is useful to evaluate the range of Faraday depth that we expect from our simulations. 
For example Fig.~\ref{fig: RM_histogram} shows the distributions of $RM$ for simulation 7 (computed using Eq.~\ref{eq: phi}, integrated over the full line-of-sight) for cubes produces with the line-of-sight perpendicular, parallel or at $45^\circ$ with the mean direction of $\vec{B}$. For each orientation the range of $RM$ values is centered on specific values, illustrating the fact that the strength of $B_\parallel$ is different in each cases. When the line-of-sight is perpendicular to $\vec{B}$, the $RM$ values are very close to zero as expected. For the case where the mean magnetic field is aligned with the LOS $\langle RM_\parallel\rangle = 5.5~\textrm{rad~m}^{-2}$. In the intermediate case, where the simulation cubes are produced with a line-of-sight at $45^\circ$ with the mean direction of $\vec{B}$, the $RM$ distribution is intermediate, with a mean $RM$ of $4.8~\textrm{rad~m}^{-2}$. In the three projections the dispersion of $RM$ is only about 1-2\,rad\,m$^{-2}$. 
These $RM$ values fall within the range observed with LOFAR in the 3C196 field \citep{Jelic2018, Bracco2020,Erceg2022}. They are also similar to what is reported by \cite{Hutschenreuter2022} in the 3C196 field, with most $RM$ values ranging from -5 to 10\,rad\,m$^{-2}$. This is interesting as the $RM$ values from Faraday measurements on extra-galactic point sources of \cite{Hutschenreuter2022} are probing greater physical depth as they are estimated. 

\subsubsection{Mock LOFAR observation}
\label{sec: MockLOFAR}
In order to compute synchrotron Stokes parameters, we developed a Julia package called \texttt{MOOSE}\footnote{\url{https://github.com/J-Berat/MOOSE/tree/main}}, described in Sect.~\ref{app: MOOSE}. For each of the 14 simulations, we used the $\vec{B}$-field and $n_e$ cubes to compute $I_\nu$, $Q_\nu$, and $U_\nu$ based on Equations~\ref{eq: epspara} to \ref{eq: Faraday angle}. 
\texttt{MOOSE} is able to treat any frequency range but in this paper we focus on LoTSS frequencies used for the study of the 3C196-field \citep{Erceg2022}, i.e $\nu$ is in the range of $[120,167]~\textrm{MHz}$ with a frequency resolution $\delta\nu = 98~\textrm{kHz}$.
As LOFAR interferometric observations are subject to large-scale filtering, we wanted to simulate this effect. To do this, we used a Gaussian kernel $G_{\textrm{ker}}$ of standard deviation $\sigma_{\textrm{LS}}$. We applied a convolution on a given image $A_{i,j}$ as follows: 
\begin{equation}
    \Tilde{A}_{i,j} = A_{i,j} - A_{i,j} \circledast G_\textrm{ker}(\sigma_{\textrm{LS}}),
\end{equation}
where $\Tilde{A}_{i,j}$ is the image where the large scales are removed.

With $Q_\nu$ and $U_\nu$ calculated, we compute $F(\phi)$ using a Julia version of Rotation Measure-synthesis (\texttt{RM-Synthesis) }\citep{Brentjens2005} implemented in \texttt{MOOSE}. Due to the frequency range, chosen to be the one of the LoTSS survey \citep{Erceg2022}, the Faraday resolution is $1.14~\mathrm{rad~m}^{-2}$ \citep[see Eq.61 in][]{Brentjens2005}. We compute $F(\phi)$ from $\phi = -10~\mathrm{rad~m}^{-2}$ to $10~\mathrm{rad~m}^{-2}$ with steps of $0.25~\mathrm{rad~m}^{-2}$ to include the range of values of $RM$ found in our simulations (see Fig.~\ref{fig: RM_histogram}).

Fig.~\ref{fig: Fphi-Tb_instrument}-left shows an example of polarized intensity at a specific Faraday depth $F(\phi)$ for simulation 7 in the case where the mean $\vec{B}$ field is along the line-of-sight. For all our simulations, $F(\phi)$ display complex structures that reflect the interplay between the electron density and magnetic field morphologies.

\subsubsection{Mock 21 cm line observation}
\label{sec: HI mock data}
To simulate 21 cm line observations of neutral hydrogen, we compute the brightness temperature $T_B(u)$ as a function of velocity $u$ along each LOS using the radiative transfer equation:
\begin{equation}
\label{eq: TbHI}
T_B(u) = \sum_k T(k) \left(1 - e^{-\tau(k,u)}\right) \cdot \exp\left(-\sum_{i<k} \tau(i,u)\right).
\end{equation}
Here the index $l$ runs over the cells along the LOS, $\tau(l,u)$ is the 21 cm optical depth at position $l$ and velocity $u$, and $T(l)$ is the gas temperature (taken as a proxy for the spin temperature).
In this formulation the term $T(l) \left(1 - e^{-\tau(l,u)}\right)$ represents the emission from cell $l$ and velocity $u$, while the exponential factor $\exp\left(-\sum_{i<l} \tau(i,u)\right)$ describes the attenuation of this emission by the gas located between position $k$ and the observer.

We assumed that the opacity $\tau(l,u)$ is 
\begin{equation}
\tau(l,u) = \frac{N_H(l,u)}{C\,T_{s}(l)}
\end{equation}
where $C = 1.82243 \times 10^{18}\,$cm$^{-2}\,$K$^{-1}\,$(km/s)$^{-1}$ and $N_H(k,u)$ is the column density in cell $l$ expressed as a Gaussian with thermal broadening, centered on the local velocity $u=v(l)$ and normalized by the local density $n_{H}(l)$:
\begin{equation}
N_H(l,u) = \frac{n_H(l)}{\sqrt{2\pi}\Delta} \times \exp \left( - \frac{(u-v(l))^2}{2 (\Delta(l))^2}\right) \, dl
\end{equation}
with
\begin{equation}
\Delta(l)= = \sqrt{\frac{k\,T(l)}{\mu\,m}}.
\end{equation}
In theory $\Delta$ should also include the turbulent velocity dispersion at scales smaller than the cell scale. We make this simplification as the typical turbulent velocity dispersion at the scale of a cell ($0.2\,$pc) is $\sim 0.6\,\mathrm{km\,s^{-1}}$, assuming a typical value of $\sigma_{u,\text{1pc}}=1\,\mathrm{km\,s^{-1}}$ and using Eq.~\ref{eq:K41}. This would become significant only for gas at $T<45\,$K which is very rare in the simulations used here.  

We produced synthetic 21 cm observations for each simulation using this formalism. 
An example of a 21\,cm channel maps is shown in Fig.~\ref{fig: Fphi-Tb_instrument}-right. Like for $F(\phi)$, the 21\,cm channel maps show complex multi-scale structure with several narrow and twisty filaments. On the other hand, contrary to $F(\phi)$ where the structure is very different depending on the orientation of $\vec{B}$ with respect to the line-of-sight, the structure in $T_B(u)$ is morphologically very similar in all directions. 

\subsubsection{\ion{H}{I} phases cubes}

\label{sec : Decomposition of the phases in the simulations}
\citet{Bracco2020} used \texttt{ROHSA} to decompose the EBHIS 21\,cm observation of the 3C196 field into phases of the neutral hydrogen. They defined the CNM as Gaussian components with $\sigma \leq 3~\textrm{km~s}^{-1}$, and WNM with components with $\sigma \geq 6~\textrm{km~s}^{-1}$. The LNM is everything in between.
To avoid having to run \texttt{ROHSA} on every 21\,cm synthetic observations of our simulation set, we chose to segment the \ion{H}{I} phases based on the 3D temperature field directly. Assuming a Mach number of 1 and 2 for the WNM and CNM respectively \citep{Marchal2024,Heiles2003}, these $\sigma_u$ thresholds correspond to the following temperature ranges: $T_\textrm{CNM} \leq 500~\textrm{K}$, $500 < T_\textrm{LNM} < 3500~\textrm{K}$ and $T_\textrm{WNM} \geq 3500~\textrm{K}$. 

In practice for a given phase (CNM, LNM, WNM) we identified the voxels in 3D corresponding to its temperature range, then produced 21\,cm synthetic observations using only those voxels. Therefore, for each simulation setup we produced four 21\,cm cubes for total \ion{H}{I}, CNM, LNM and WNM.

\section{Results}
\label{sec: Results}
\subsection{Global properties of $T_B(u)$ and $F(\phi)$}
\label{sec: Visual results}

\begin{figure*}[!ht]
\centering
\includegraphics[width=\linewidth]{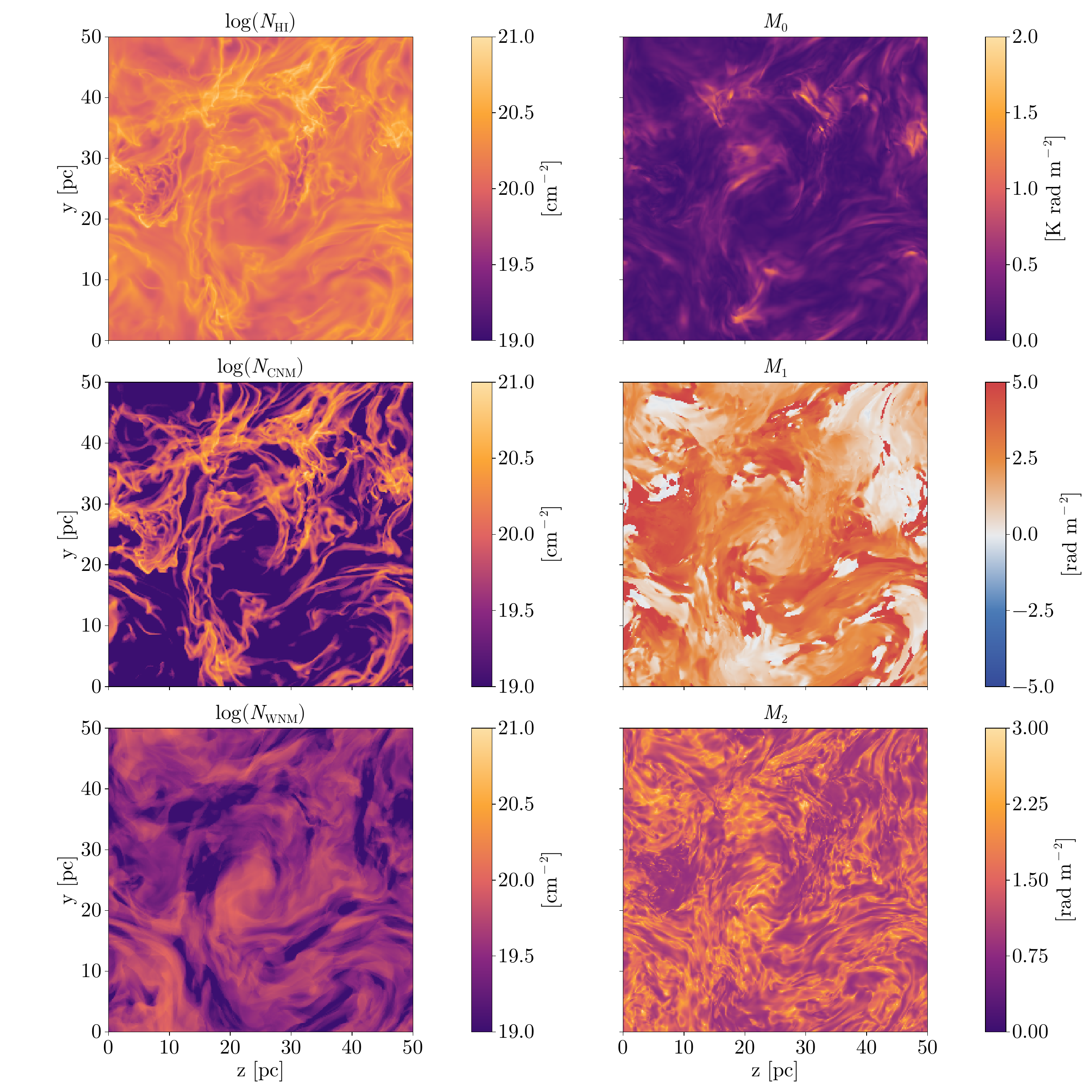}
    \caption{Column density (\textit{left panels}) and Faraday moment maps (\textit{right panels}) for case-study simulation, with integration along the mean magnetic field. \textit{Top to bottom}: logarithmic column density maps of total \ion{H}{I}, CNM, and WNM; moments 0, 1 and 2 of the Faraday spectrum.}
    \label{fig: NmomentsX}
\end{figure*}

\begin{figure*}[!ht]
\centering
\includegraphics[width=\linewidth]{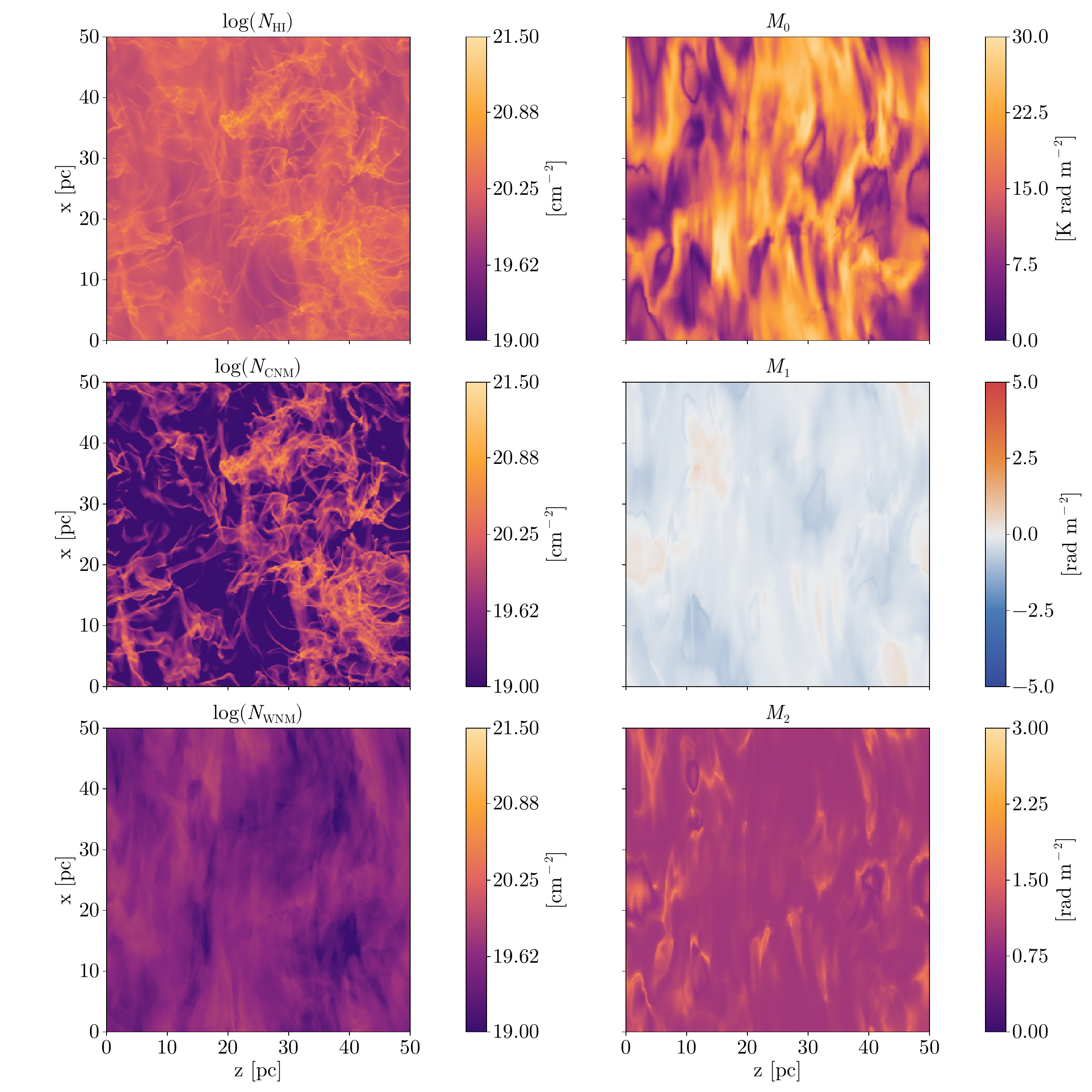}
    \caption{Same as Fig.~\ref{fig: NmomentsX}, but for the case where the LOS is perpendicular to the mean magnetic field. Due to the weak line-of-sight magnetic component, $M_1$ values remain close to zero and $M_2$ is narrower across the field. While the density structures of CNM and WNM are similar to the parallel case, the Faraday moments show significantly reduced signal, consistent with expectations for a transverse magnetic field configuration.}
    \label{fig: NmomentsY}
\end{figure*}

The \ion{H}{I} and Faraday structures of simulation 7 are illustrated in Figures~\ref{fig: NmomentsX} and \ref{fig: NmomentsY} with views parallel and perpendicular to the mean $\vec{B}$ field. On the left, these figures show the column density maps of the \ion{H}{I} components (total, CNM and WNM) highlighting the striking filaments of the CNM as well as the diffuse structure of the more volume filling WNM. On the right are shown moments ($M_0$, $M_1$ and $M_2$) of the Faraday cube - see Sect.~\ref{sec: Moments and effective width}.
These can be compared directly to the moments of the 3C196-field LoTSS observations (Fig.~\ref{fig: moments_3C196}).

\begin{figure*}[!ht]
\centering
\includegraphics[width=0.45\linewidth]{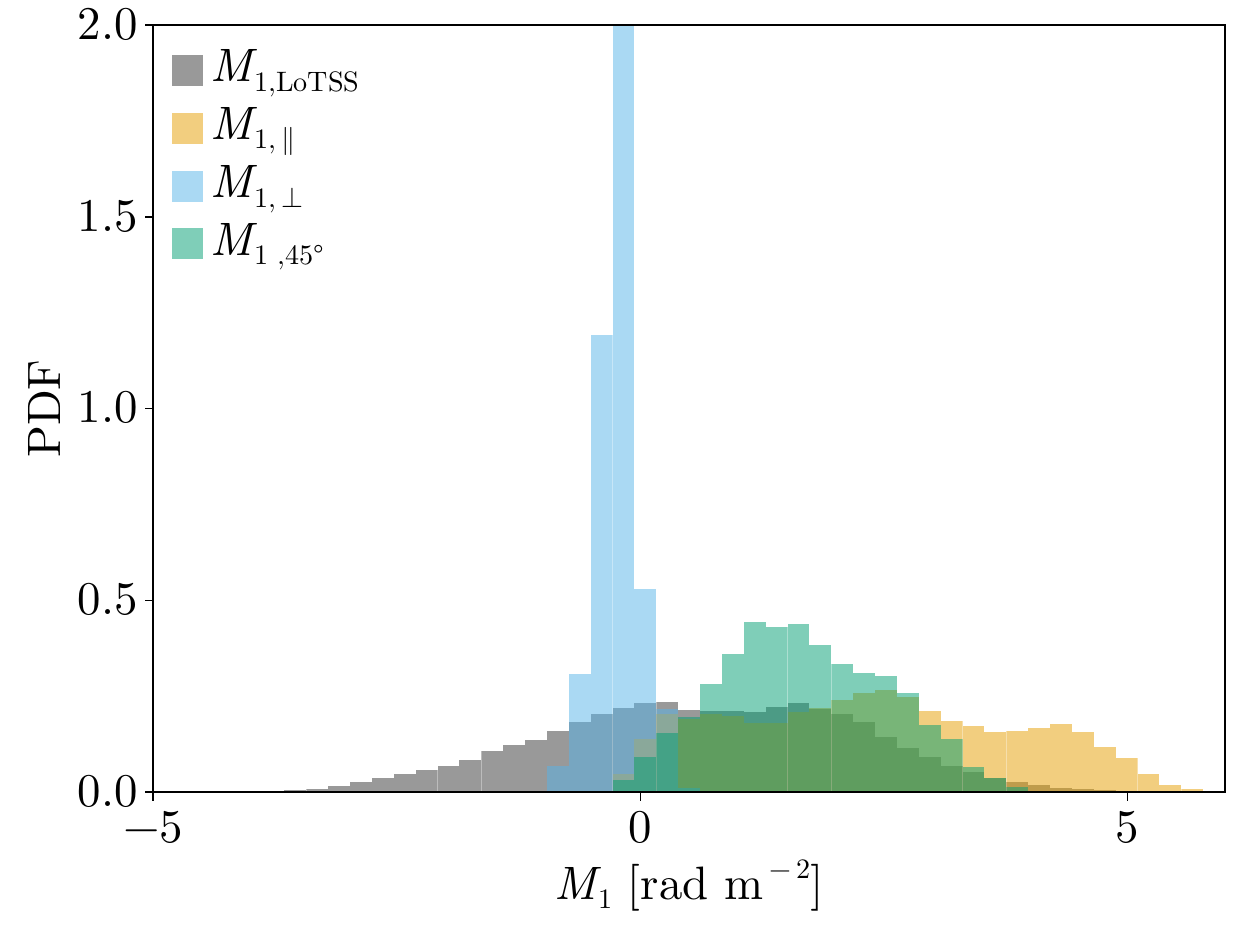}
\includegraphics[width=0.45\linewidth]{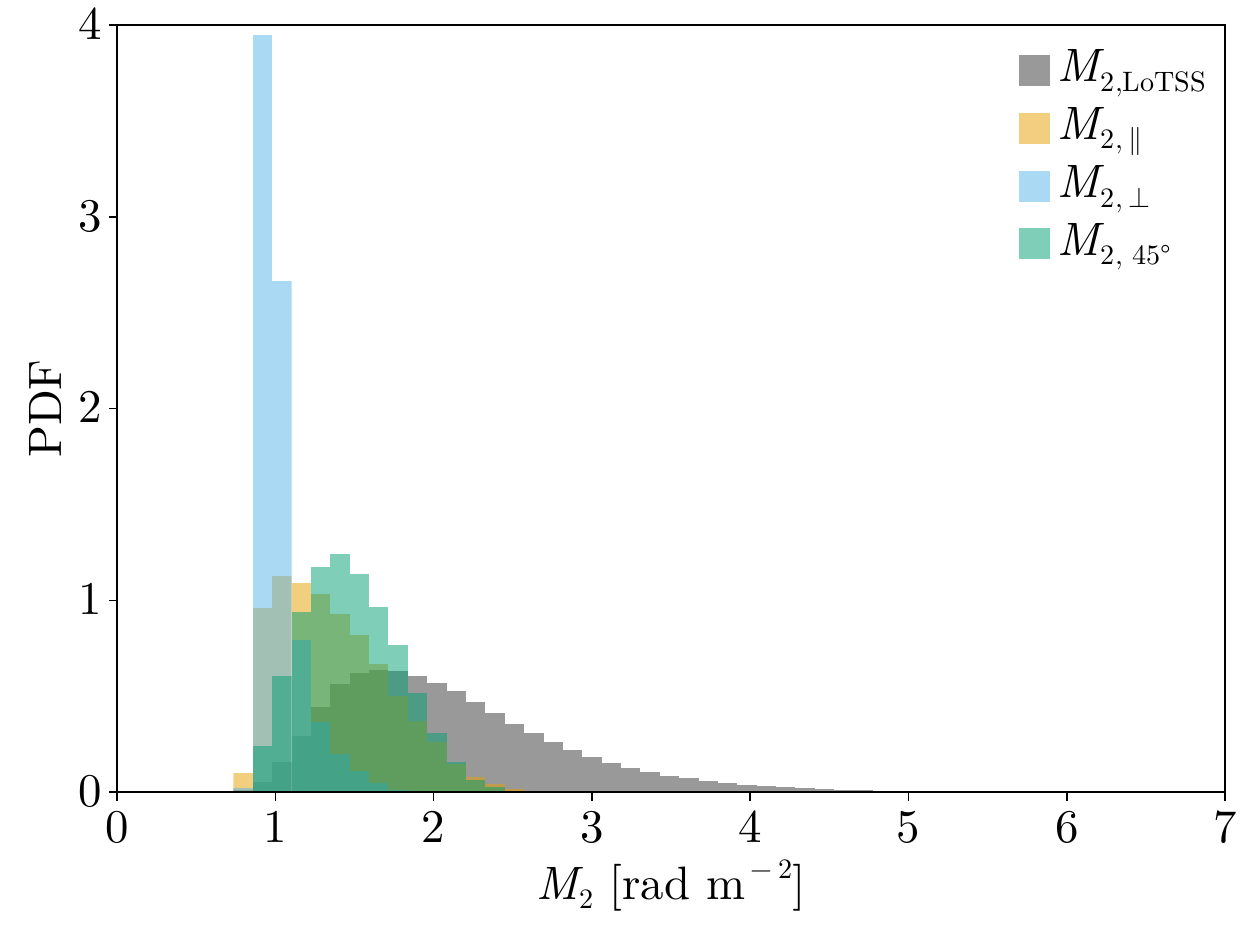}
\caption{Probability density functions of Faraday spectral moments for the 3C196 field (gray) and for the case-study simulation with three magnetic field orientations: parallel (yellow), perpendicular (blue), and $45^\circ$ rotated (green). \textit{Left:} First-order moment $M_1$, showing higher values in the parallel case and near-zero mean in the perpendicular case. \textit{Right:} Second order moment $M_2$ of the Faraday spectrum.}
\label{fig: histogram_M1 and M_2}
\end{figure*}

\begin{figure*}[!ht]
\centering
\includegraphics[width=0.9\linewidth]{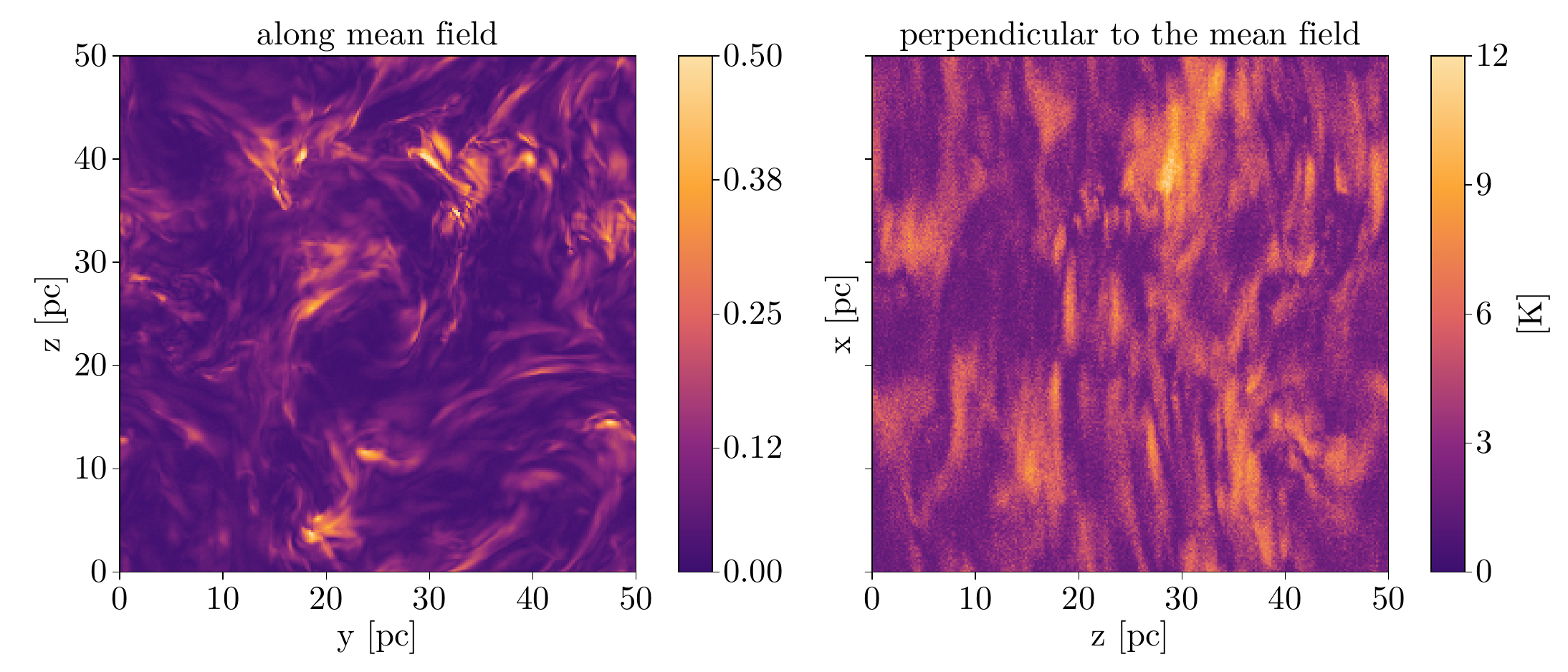}
    \caption{Maximum polarized intensity maps ($P_{\max}$) from the case-study simulation, illustrating the presence and morphology of depolarization canals. \textit{Left:} integration along the LOS parallel to the mean magnetic field. \textit{Right:} integration perpendicular to the mean field. Linear canal-like features are prominent in the parallel case, while more circular structures dominate in the perpendicular configuration.}
    \label{fig: Pmax}
\end{figure*}

The Faraday moments of the simulations (Figures~\ref{fig: NmomentsX} and \ref{fig: NmomentsY}) are strongly influenced by the orientation of the magnetic field. In the case perpendicular to the mean field, the values of $M_1$ remain close to zero due to the weak component of field along the LOS. On the other hand, when the magnetic field is parallel to the LOS we observe values of $M_1$ that are consistently positive, falling within the positive range observed in the $M_1$ map of \citet{Erceg2022}. 

Regarding the spread of the Faraday spectra estimated using $M_2$, the values are very small in the case where $\vec{B}$ is perpendicular to the LOS, very close to theoretical RMSF of LoTSS which is $1.03 \sim \textrm{rad m}^{-2}$ estimated for the theoretical RMSF of LoTSS. 
On the other hand, the case where $\vec{B}$ is parallel to the LOS shows $M_2$ values comparable to the ones estimated in the LoTSS 3C196 field.

The histograms of the first-order moment $M_1$ are shown in Fig.~\ref{fig: histogram_M1 and M_2} (top panel). For the 3C196 field, $M_1$ has a mean value of approximately $0.2~\mathrm{rad~m^{-2}}$ with a standard deviation of $0.9~\mathrm{rad~m^{-2}}$. Among the simulations, the perpendicular case provides the closest match in terms of mean value, which reaches $\langle M_1 \rangle  = -0.2~\mathrm{rad~m^{-2}}$, while the parallel and $45^\circ$ cases have $\langle M_1 \rangle  = 2.4~\mathrm{rad~m^{-2}}$ and $\langle M_1 \rangle  = 1.7~\mathrm{rad~m^{-2}}$ respectively.  

The Faraday spread $M_2$ (de-biased from noise) measured in the 3C196 field has a median value of approximately $1.7~\mathrm{rad~m^{-2}}$ (see right panel of Fig.~\ref{fig: histogram_M1 and M_2}). Observed values, even after de-biasing from noise, are generally slightly higher than what is recovered in the synthetic observations constructed without noise. The best agreement is found when the emission is integrated along the direction rotated by $45^\circ$ to the mean magnetic field, yielding a mean of $\langle M_2\rangle = 1.5~\mathrm{rad~m^{-2}}$. This is close to the the case where $B$ is parallel to the line-of-sight ($\langle M_2\rangle = 1.4~\mathrm{rad~m^{-2}}$). but a narrower distribution. The perpendicular configuration shows lower values as expected because the $\vec{B}$ component along this direction is very low with $\langle M_2 \rangle =1.0~\mathrm{rad~m^{-2}}$. 

One striking features of the LoTSS polarization data shown in  \cite{Erceg2022} is the presence of long and thin depolarization canals, i.e. filamentary regions with the maximum of $F(\phi)$, $P_\mathrm{max}$, close to zero (for more details, see \citet{sokoloff98, Haverkorn2004}). Maps of $P_\mathrm{max}$ (LOS perpendicular and parallel to $\vec{B}$) are shown in Fig.~\ref{fig: Pmax}. Structures with almost zero intensity are seen in the simulations, especially in the case where the mean $\vec{B}$ field is aligned with the LOS.
However, the long, straight canals detected in the LoTSS survey are not reproduced. 

\subsection{Correlation between LoTSS and EBHIS data}
\label{sec: HOG HI/Faraday observations}
Following the same methodology of \citet{Bracco2022}, we compare cube of \ion{H}{I} brightness temperature cubes with Faraday cubes using the Histogram of Oriented Gradients method \citep{Soler19}. The core principle of HOG is to measure the morphological alignment between two images, $A_{i,j}$ and $B_{i,j}$, where $i$ and $j$ denote the image coordinates. It relies on the assumption that the local appearance and shape of an image can be fully characterized by the distribution of its local intensity gradients or edge directions.  
From this assumption, the gradients of the two images can be computed using Gaussian derivatives, yielding $\nabla A_{i,j}$ and $\nabla B_{i,j}$. Following the definition in \citet{Soler19}, the angle between the two gradient vectors, $\alpha$, can then be determined as follows:

\begin{equation}
\label{eq: alpha_i,j}
\alpha_{i,j} = \arctan{\left(\frac{(\nabla A_{i,j} \times \nabla B_{i,j}) \cdot \hat{z}}{\nabla A_{i,j} \cdot \nabla B_{i,j}}\right)},
\end{equation}
where $\hat{z}$ is the unit vector perpendicular to the plane of sky. HOG tests whether the distribution of angle $\alpha$ is uniformly distributed or has a preferential alignment at $\alpha = 0$, using the projected Rayleigh statistics defines as
\begin{equation}
\label{eq: V}
V = \frac{\sum\limits_{i,j} \omega_{i,j} \cos{(2\alpha_{i,j})}}{\sum\limits_{i,j} (\omega_{i,j}^2/2)^{1/2}},
\end{equation}
where $\omega_{i,j}$ is the statistical weight corresponding to each $\alpha_{i,j}$. The Rayleigh statistics parameter $V$ is thus the estimate of how well the two images are morphologically correlated. For more details about this technique see \citet{Soler19}.

The observational maps do not have the same dimensions as those from the simulations. To address this, we introduce a normalization factor, $V_{\textrm{max}}$, defined as the maximum value of $V$.  
The maximum of $V$ is attained when the gradients of the two maps are perfectly aligned (i.e., $\alpha_{i,j} = 0$) and supposing the weights are all the same (i.e., $\omega_{i,j} = \omega$) \citep[also detailed in][]{Mininni2025}. From Eq.~\ref{eq: V}, this yields $V_\textrm{max} = (2N)^{1/2}$.

Specifically, let $F_{i,j,\phi}$ represent the Faraday cube and $T_{b_{i,j,u}}$ the 21 cm brightness temperature cube, where $i, j$ are  pixel indices on the plane of the sky.
For each pair $(\phi, u)$, we apply HOG using a local kernel around $F_{i, j}$ and $T_{b_{i,j}}$. This allows us to construct maps parameterized by $(\phi, u)$, illustrating the spatial correlation between specific $F(\phi)$ and corresponding $T_b(u)$.

\begin{figure*}
\centering\includegraphics[width=0.85\linewidth]{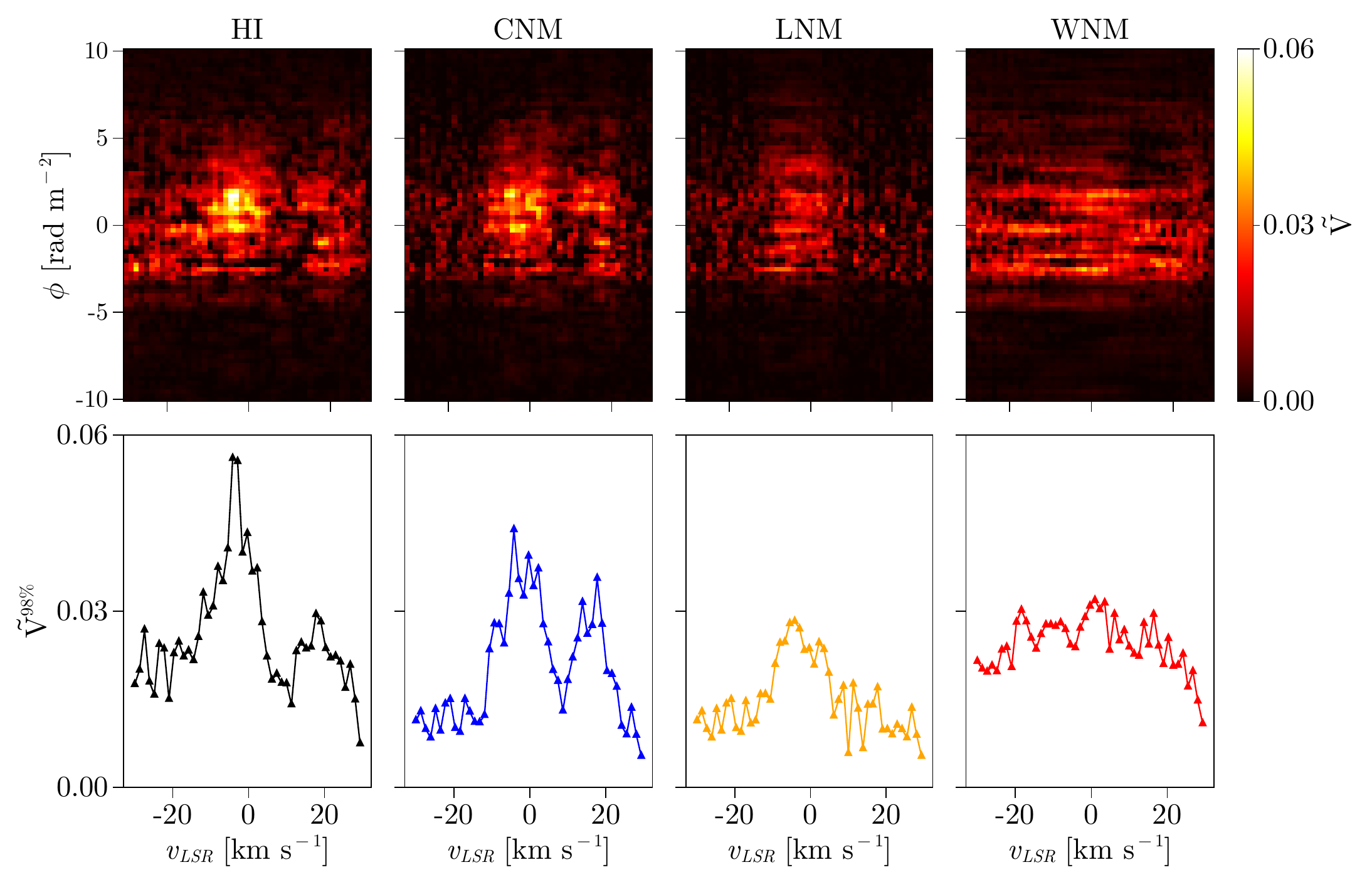}
     \caption{HOG $\Tilde{V}$ for the 3C196-field LoTSS observation. \textit{Upper panels}: 
     Map of $\Tilde{V}$ illustrating the spatial correlation between each pairs of channel maps ($T_B(u)$, $F(\phi)$) for all values of $u$ and $\phi$.
    \textit{Bottom panels}: $\Tilde{V}^{98\%}$, the $98^{\text{th}}$ percentile value of $\Tilde{V}$ along the $\phi$ axis.}
\label{fig: VoverVmax LOFAR}
\end{figure*}

The correlation in the 3C196 field reported in \citet{Bracco2020} has now been updated using LoTSS data, applying the same phase decomposition with \texttt{ROHSA} from EBHIS data. 
In Fig.~\ref{fig: VoverVmax LOFAR}, the ratio $\Tilde{V} = V/V_\textrm{max}$ is represented on a map with axes $\phi$ and $v_\textrm{LSR}$. In the total \ion{H}{I} and CNM maps, two distinct velocity regions can be identified: one around $v_\textrm{LSR} = 0~\textrm{km~s}^{-1}$ and another around $v_\textrm{LSR} = 20~\textrm{km~s}^{-1}$, similar to the findings of \citet{Bracco2020}.  
Along the $\phi$-axis, a broad structure is observed, spanning approximately $5~\textrm{rad~m}^{-2}$, with a central peak around $\phi = 0~\textrm{rad~m}^{-2}$.
This structure appears bright across all velocities in \ion{H}{I} and WNM. For the CNM, velocities extend from $v_\textrm{LSR} = -10~\textrm{km~s}^{-1}$ to $v_\textrm{LSR} = 20~\textrm{km~s}^{-1}$, with a strong correlation at $20~\textrm{km~s}^{-1}$. In contrast, for the LNM phase, this structure span a more limited velocity range, from $v_\textrm{LSR} = -20~\textrm{km~s}^{-1}$ to $v_\textrm{LSR} = 0~\textrm{km~s}^{-1}$.

The highest value is $\Tilde{V} \approx 0.06$, but the maximum being sensitive to outliers we show the $98^\mathrm{th}$ percentile of $\Tilde{V}$, $\Tilde{V}^{98\%}$, for each $\phi$ on the lower panels of Fig.~\ref{fig: VoverVmax LOFAR}. 
Based on this HOG metric, the strongest correlation between $T_B(u)$ and $F(\phi)$ is found for the CNM (also seen in the total \ion{H}{I}) compared with the WNM and LNM components. This is compatible with the results of \cite{Bracco2020}.

In order to further quantify the level of correlation of the CNM and WNM phases, we used the maximum of $\Tilde{V}^{98\%}$ for each phase, and computed the ratio: 
\begin{equation}
    \eta = \frac{\max({\Tilde{V}_\textrm{WNM}^{98\%}})}{\max({\Tilde{V}_\textrm{CNM}^{98\%}})}.
    \label{eq: eta}
\end{equation}
This value reflects the strength of the dominant correlation among those two: $\eta < 1$ is indicative of a stronger correlation of the Faraday data with the CNM, while $\eta > 1$ shows a stronger correlation with the WNM. 

In the 3C196 field of view, we find $\eta = 0.73$, which supports the visual observation in Fig.~\ref{fig: VoverVmax LOFAR} that the CNM exhibits the strongest correlation with the Faraday data.

\subsection{Correlation between Faraday tomography and \ion{H}{I} emission with numerical simulations}
\label{sec: HOG HI/Faraday}

\begin{figure*}[!ht]
\centering
\includegraphics[width=0.85\linewidth]{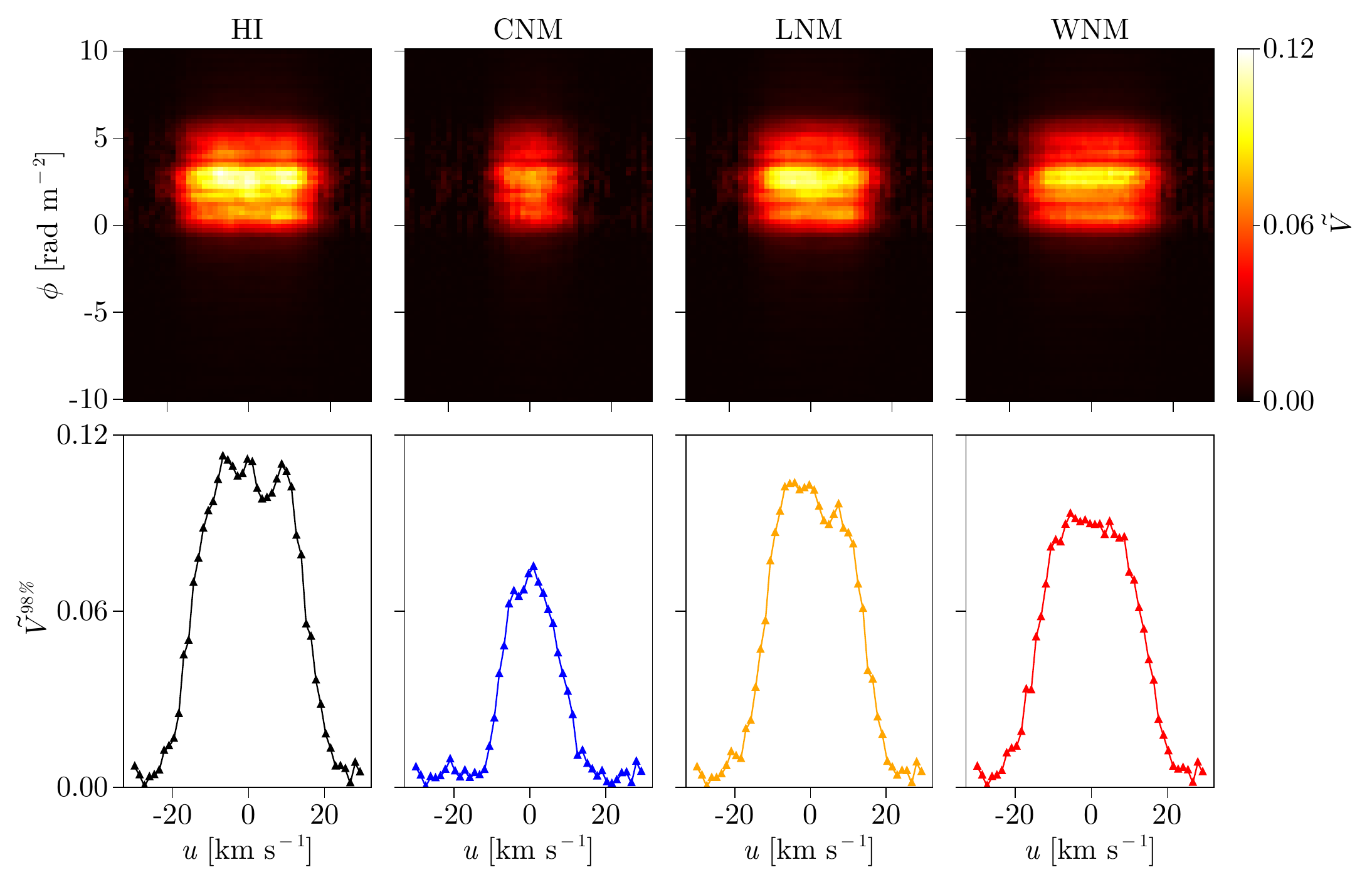}
    \caption{HOG $\Tilde{V}$ for simulation 7 orientated along the mean magnetic field. \textit{Top  panels}: Map of $\Tilde{V}$ illustrating the spatial correlation between each pairs of channel maps ($T_B(u)$, $F(\phi)$) for all values of $u$ and $\phi$. We considered the cases where $T_B(u)$ is computed with the full \ion{H}{I} or using subset of pixels representing the CNM, LNM and WNM gas (see Sect.~\ref{sec : Decomposition of the phases in the simulations}). \textit{Bottom panels}: $\Tilde{V}^{98\%}$, the $98^{\text{th}}$ percentile value of $\Tilde{V}$ along the $\phi$ axis.
}
    \label{fig: HOG_7 along mean field}
\end{figure*}

\begin{figure*}[!ht]
\centering
\includegraphics[width=0.85\linewidth]{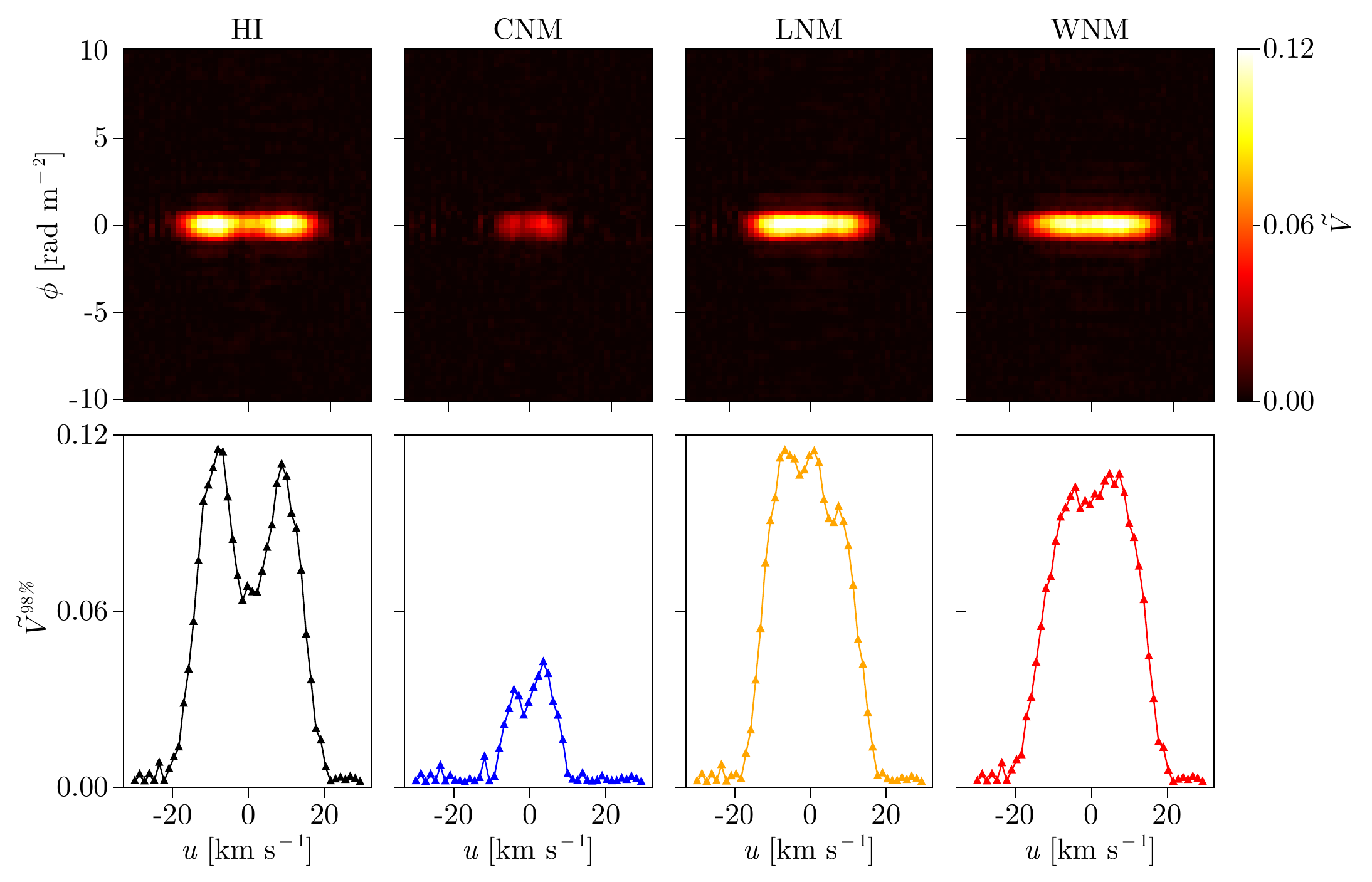}
    \caption{HOG $\Tilde{V}$ for simulation 7 orientated perpendicular to the mean magnetic field. See Fig.~\ref{fig: HOG_7 along mean field} for details.
}
    \label{fig: HOG_7 perp mean field}
\end{figure*}

We applied the same HOG methodology on the synthetic observations of our simulation set. Figures~\ref{fig: HOG_7 along mean field} and \ref{fig: HOG_7 perp mean field} present the HOG results for the correlation of the \ion{H}{I} phases with the Faraday cube for simulation 7, for two orientations (mean $\vec{B}$ parallel and perpendicular to the LOS). For both orientations the WNM and LNM exhibit higher correlation with Faraday structures than the CNM, in contrast with the results on the 3C196 field of view. 

For the case where the mean $\vec{B}$ field is aligned with the LOS (Fig.~\ref{fig: HOG_7 along mean field}), the WNM and LNM have $\Tilde{V}$ values peaking at $0.12$. The CNM, although less correlated, reaches values ($\Tilde{V} \sim 0.07$) very close to the maximum $\Tilde{V}$ observed in the 3C196 field. The curve of $\Tilde{V}^{98\%}(\phi)$ confirms this trend, with the WNM displaying the broadest and strongest peak, followed by the LNM and CNM. The strongest correlation is observed around $\phi \approx 3~\textrm{rad~m}^{-2}$, consistent with the expected Faraday depth range for a magnetic field aligned with the LOS.

For the case where the magnetic field lies in the plane of the sky (Fig.~\ref{fig: HOG_7 perp mean field}), the values of $\Tilde{V}$ are similar but the structure in the map of $\Tilde{V}$ is much narrower in $\phi$ space compared to the case $\vec{B}$ parallel to LOS. The WNM and LNM still exhibit the highest correlation, but here the CNM correlation is even lower. In this configuration, the correlation is concentrated near $\phi \approx 0~\textrm{rad~m}^{-2}$, which is expected since a predominantly perpendicular field results in minimal Faraday rotation.

Overall, the case-study simulation produces $\Tilde{V}^{98\%}$ values of total \ion{H}{I}, WNM and LNM that are approximately twice as high as those observed in the 3C196 field. The CNM phase has $\Tilde{V}$ values lower than the other phases but they reach levels consistent with observational values ($\Tilde{V}\approx 0.06$).

Interestingly, for the case where $\vec{B}$ is perpendicular to the LOS (Fig.~\ref{fig: HOG_7 perp mean field}), the total \ion{H}{I} displays a $\Tilde{V}^{98\%}(u)$ curve with a distinctive double-peaked U-shaped profile. 
As shown in Fig.~\ref{fig: LIC along and perp}-right, 
in this orientation some dense structures associated with the CNM appear misaligned or even orthogonal to the projected $\vec{B}$ field. This morphological decoupling reduces the coherence between cold \ion{H}{I} structures and the Faraday tomographic data. The result is a decrease of $\Tilde{V}$ in velocity channels dominated by CNM emission, inducing this U-shaped form in $\Tilde{V}^{98\%}(u)$.

\begin{figure*}
\centering
\includegraphics[width=0.95\linewidth]{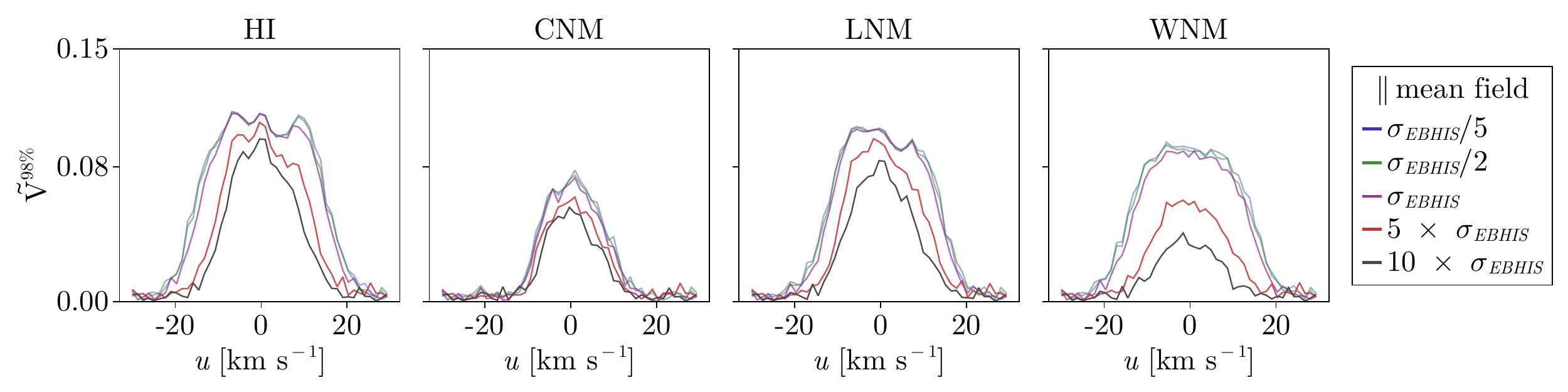}
    \caption{Effect of increasing noise in the \ion{H}{I} data on the HOG correlation with Faraday tomography structures for the case-study simulation along the mean field. Each panel displays the 98th percentile of the normalized Rayleigh statistic $\tilde{V}^{98\%}$ as a function of Faraday depth $\phi$, for the CNM and WNM phases. The different lines correspond to different levels of added Gaussian noise in the \ion{H}{I} cubes, ranging from $\sigma_{\mathrm{EBHIS}}/5$ to $10 \times \sigma_{\mathrm{EBHIS}}$. While increasing noise decreases the overall correlation strength, the WNM remains more affected than the CNM, particularly at high velocities.}
    \label{fig: simu 7 HOG noise HI}
\end{figure*}

\begin{figure*}
\centering
\includegraphics[width=0.95\linewidth]{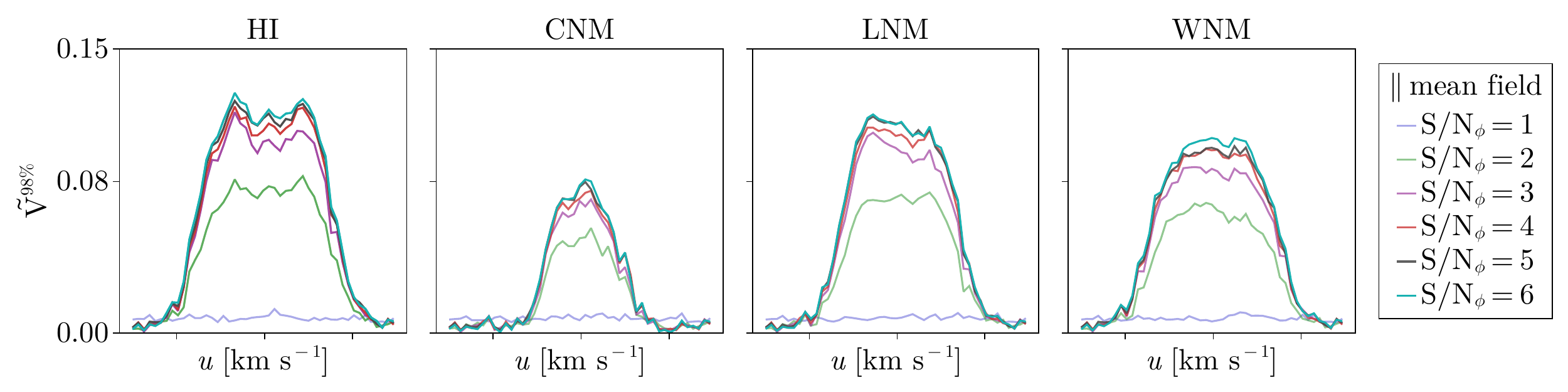}
    \caption{Impact of increasing noise in the Faraday data on the HOG correlation between phase-separated \ion{H}{I} structures and Faraday depth features for the case-study simulation along the mean field. Each panel shows the 98th percentile of the normalized Rayleigh statistic $(\tilde{V}^{98\%}$ as a function of Faraday depth $\phi$, for the CNM and WNM phases. The panels correspond to increasing signal-to-noise ratios (S/N) in Faraday space: from 1 to 6, as indicated. Higher S/N enhances the correlation amplitude for both phases, but the relative dominance of the WNM is preserved across all regimes.}
    \label{fig: simu 7 HOG noise QU}
\end{figure*}

\subsection{Noise effects on HOG results}
\label{sec: Noise effects on HOG results}

To assess the impact of noise on the correlation between the \ion{H}{I} phases and the Faraday structures, we added white noise into the mock \ion{H}{I} and Faraday data. For \ion{H}{I}, noise was added in each velocity channel at various levels with respect to the $\sigma_{\textrm{EBHIS}}=90~\textrm{mK}$ noise \citep{winkel2016}: $\sigma_\mathrm{EBHIS}/5$, $\sigma_\mathrm{EBHIS}/2$, $\sigma_\mathrm{EBHIS}$, $2 \times \sigma_\mathrm{EBHIS}$, and $10 \times \sigma_\mathrm{EBHIS}$. 
For Faraday, the noise was added to $Q_\nu$ and $U_\nu$, with a noise amplitude tuned to reproduce specific S/N ratios in Faraday depth space, including the S/N of the observational data which is 5, as well as both lower and higher S/N regimes.

As shown in Fig.~\ref{fig: simu 7 HOG noise HI}, the addition of noise to the \ion{H}{I} data predominantly affects the WNM structures. Specifically, when increasing the noise level from $\sigma_\mathrm{EBHIS}/5$ to $10 \times \sigma_\mathrm{EBHIS}$, the maximum correlation value for the CNM decreases by a factor of approximately 1.5, whereas for the WNM, the corresponding decrease reaches a factor of 2.5. 

Two effects are at play here. First the WNM has a broad emission line due to its larger thermal broadening. Therefore a given column density of WNM is more spread out in velocity with lower values of $T_B(u)$ than the same column density of CNM that will be narrow and brighter. Second, the WNM has an intrinsic diffuseness and lower spatial contrast compared to the CNM which forms narrow, high-contrast filaments. As a result, the WNM large scale structure with smoother gradients are more easily suppressed by noise, particularly at high spatial frequencies where gradient-based methods like HOG are most sensitive.

Contrary to $T_B(u)$, adding noise to $Q_\nu$ and $U_\nu$ affects both the WNM and CNM in a similar manner. As shown in Fig.~\ref{fig: simu 7 HOG noise QU}, increasing the signal-to-noise ratio in the Faraday space ($S/N_\phi$) from 2 to 6 leads to an increase in the maximum correlation value by a factor of approximately 1.5 for both phases. In the case where the signal amplitude equals that of the noise (i.e., $S/N_\phi$ = 1), the resulting correlation between the Faraday cube and the \ion{H}{I} data is effectively random ($\Tilde{V}^\textrm{98\%} \approx 0$), as expected. The case perpendicular to the mean field gives similar results.

{In conclusion, noise in $T_B(u)$ tends to decrease $\eta$ (i.e., the noise decreases the correlation of WNM compared to CNM) while noise in $F(\phi)$ does not affect $\eta$ has it reduces the absolute value of $\Tilde{V}$ for all phases. }

\subsection{Kernel size effect on HOG results}
\label{Kernel size effect on HOG results}

\begin{figure*}[!ht]
\sidecaption
\includegraphics[width=12cm]{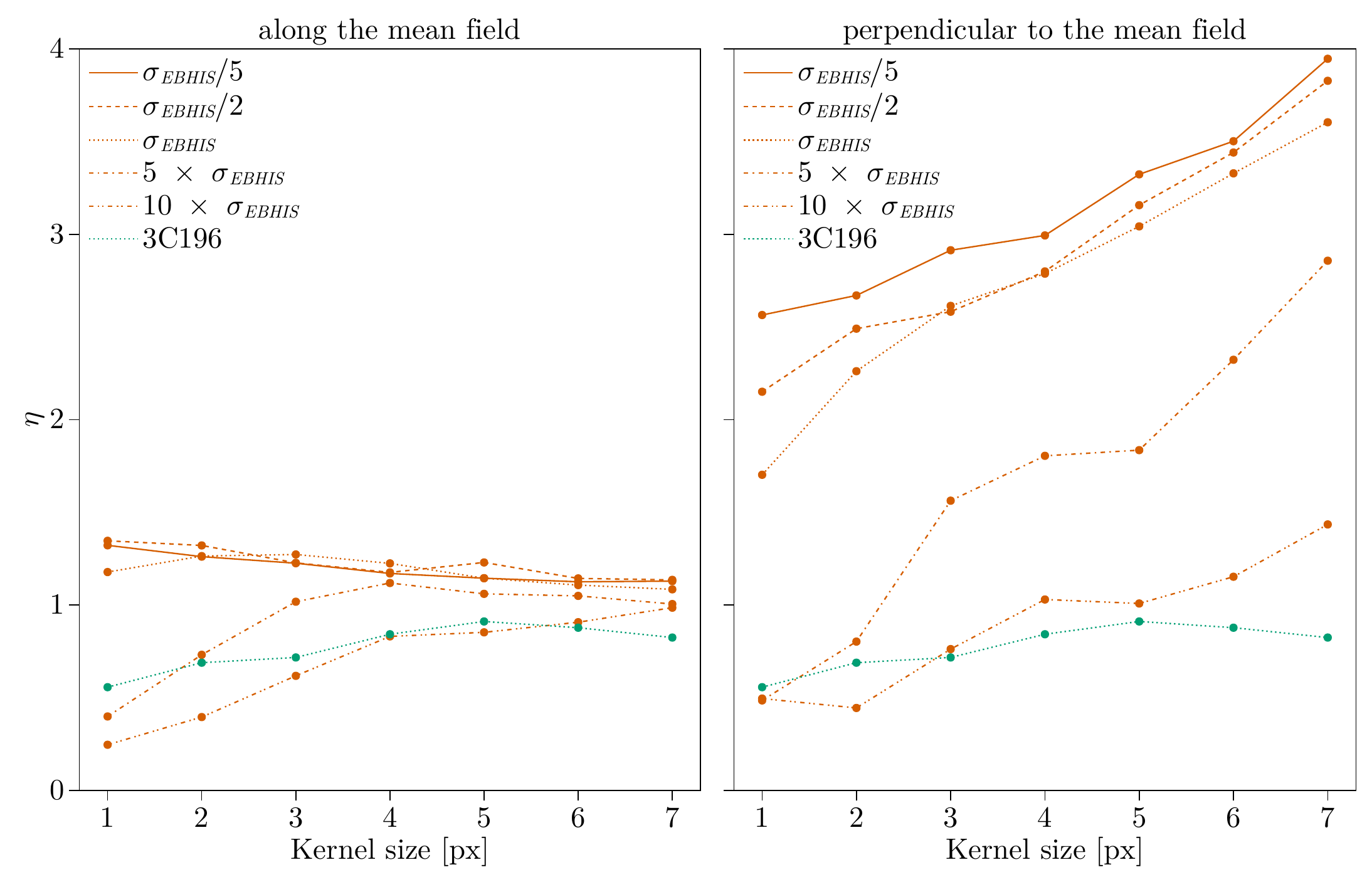}
    \caption{Dependence of the HOG correlation maxima ratio $\eta$ on the derivative kernel size used in the gradient computation for different levels of Gaussian noise added to the \ion{H}{I} data. Left: Case-study simulation with the magnetic field aligned with the LOS. Right: same simulation with the mean field perpendicular to the LOS. The observational value from the 3C196 field is shown for comparison (dashed line).}
    \label{fig: eta ratio}
\end{figure*}

An important parameter of the HOG method is the derivative kernel size, $\sigma_{\textrm{HOG}}$, that dictates the scale over which the spatial gradient is calculated. This parameter might have a different effect on $\Tilde{V}$ for the WNM and CNM as they are not structured at the same scale;
CNM structures exhibit a more small-scale filamentary morphology, whereas the WNM is present across a broad range of scales with limited small scale contrast. 
In this section we evaluate the effect of 
$\sigma_{\textrm{HOG}}$ on $\eta$. We do so for different values of the noise level on $T_B(u)$ as it has an effect on $\eta$ (see Sect.~\ref{sec: Noise effects on HOG results}). 
Figure~\ref{fig: eta ratio} shows the variation of $\eta$ as a function of $\sigma_{\textrm{HOG}}$ (in pixels) for different \ion{H}{I} noise levels added to simulation 7, for $\vec{B}$ parallel (left) and perpendicular (right) to the LOS. 
The observational reference field 3C196 is also shown in this figure, for comparison. The values of $\eta$ for the 3C196 field remains consistently below unity across all kernel sizes $\sigma_{\textrm{HOG}}$ indicating that the CNM is systematically more correlated with the Faraday structures than the WNM.

The results on the mock data of simulation 7 shows a significant variation of $\eta$ as a function of $\sigma_{\textrm{HOG}}$, \ion{H}{I} noise level and orientation of $\vec{B}$ with respect to the LOS. 
For the case $\vec{B}$ parallel to the LOS, the value of $\eta$ is almost independent of $\sigma_{\textrm{HOG}}$ while $\eta$ increases significantly with $\sigma_{\textrm{HOG}}$ in the case where $\vec{B}$ is perpendicular to the LOS. Strong noise in the \ion{H}{I} data produces an increase of $\eta$ with $\sigma_{\textrm{HOG}}$ as the effect of the noise affecting the WNM is gradually reduced. These variations for a single simulation show that observational results on $\eta$ can depend significantly on the observational conditions (noise level, physical resolution of the observations) but also on the mean orientation of $\vec{B}$ with respect to the LOS. 

That said, our analysis is showing that the only way to reproduce the values of $\eta$ seen in the 3C196 field with simulation 7 is to add a high noise level in the \ion{H}{I} mock data in order to lower the WNM correlation at the benefit of the CNM. This is discussed further in Sect.~\ref{sec: discussion}.

\subsection{Variation with physical parameters}

\subsubsection{Turbulence strength}

The HOG analysis described in the previous sections and illustrated on simulation 7 was also performed on our full simulations set. First, in order to understand the nature of the signature observed in Figures~\ref{fig: HOG_7 along mean field} and \ref{fig: HOG_7 perp mean field}, we show the $\Tilde{V}^{98\%}(u)$ curves for all simulations in Figures~\ref{fig: paraVoverVmax_allsimu_para} and \ref{fig: paraVoverVmax_allsimu_perp}. We chose not to apply any instrument model (LOFAR beam, noise on Faraday and \ion{H}{I}) to estimate the effect of turbulence level alone.

\begin{figure*}
\sidecaption
\includegraphics[width=12cm]{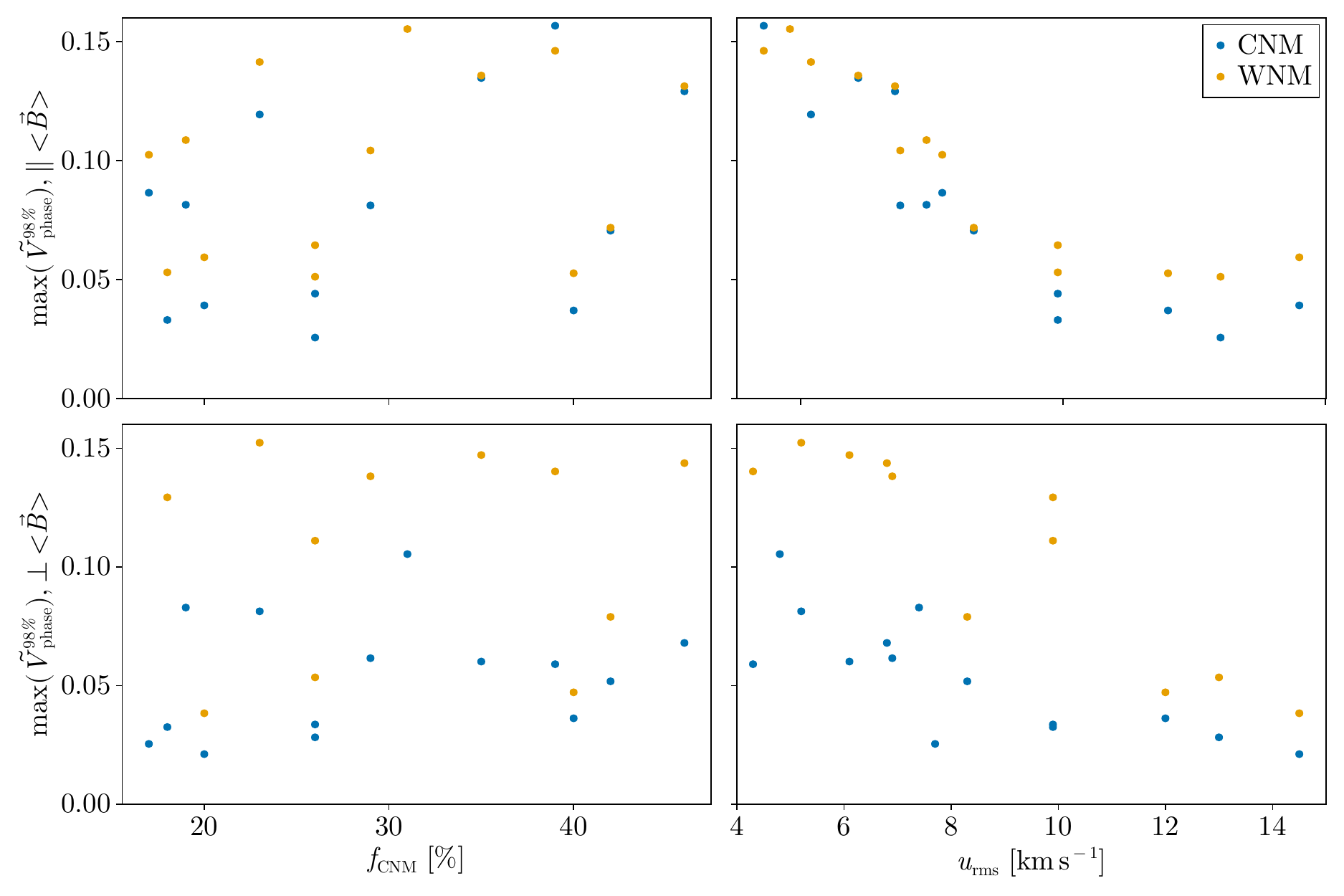}
    \caption{Compilation for all simulations of the maximum value of the 98th percentile of the normalized Rayleigh statistic $\tilde{V}^{\mathrm{98\%}}$ as a function of $f_{\mathrm{CNM}}$ (left) and $u_\textrm{RMS}$ (right). Cases for $B$ perpendicular (bottom) and parallel (top) to the line of sight are shown. In each panel, the data points show the values of $\tilde{V}^{\mathrm{98\%}}$ for the CNM (blue) and WNM (yellow). }
    \label{fig:maxV}
\end{figure*}

We observed that simulations with lower turbulence strength have higher $\Tilde{V}$ values than simulation with stronger turbulent driving, see Fig.~\ref{fig:maxV}. This suggests that lower turbulence, and therefore weaker magnetic fluctuations, enhances the correlation between magneto-ionized observables and the \ion{H}{I} phases.
However, we also note that $\eta$ is always higher than what is seen in observational data whatever the turbulence level (i.e., the WNM remains more correlated to $F(\phi)$ than the CNM). We recall that this experiment was done without any noise in the mock data and, as shown in Fig.~\ref{fig: eta ratio}, noise in the \ion{H}{I} data will tend to lower $\eta$.

\subsubsection{CNM fraction}
\label{CNM fraction impact on HOG results}

Finally, one goal of the current study was to evaluate if the fraction of the \ion{H}{I} in the CNM phase has an impact on the measured correlation between $T_B(u)$ and $F(\phi)$. 
To do so we examined how $\eta$ varies as a function of $f_{\mathrm{CNM}}$. The aim is to determine whether a higher proportion of CNM in the simulated \ion{H}{I} data systematically enhances the correlation with Faraday structures, and whether this trend depends on the orientation of the magnetic field with respect to the LOS.

\begin{figure*}[!ht]
\sidecaption
\includegraphics[width=12cm]{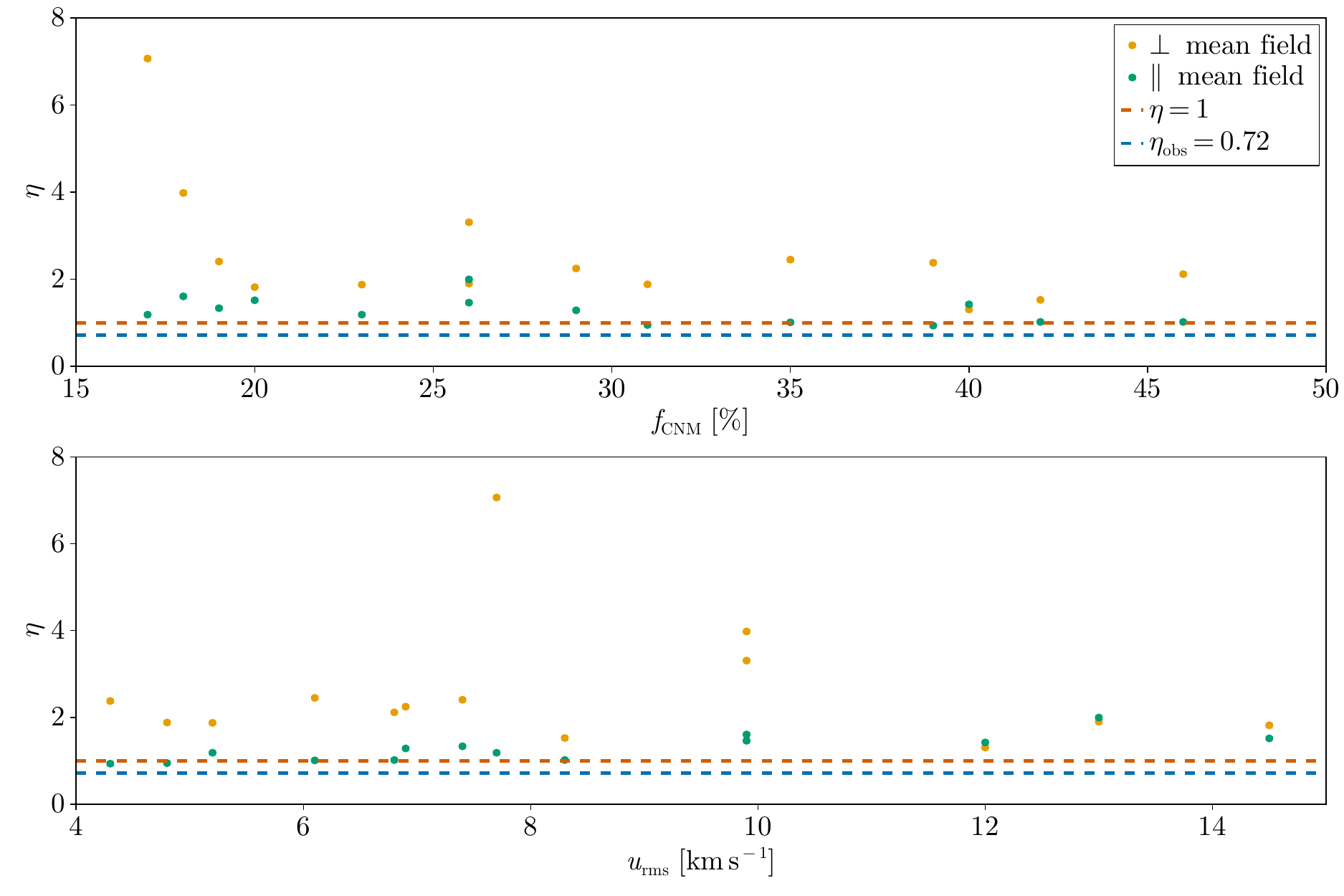}
    \caption{Compilation for all simulations of the maximum HOG correlation coefficient $\eta$ as a function of $f_{\mathrm{CNM}}$ (top) and $u_\textrm{RMS}$ (bottom). The data points show $\eta$ measured in orientations of the line of sight parallel (green) and perpendicular (yellow) to the mean magnetic field. The dashed line at $\eta = 1$ separates CNM-dominated ($\eta < 1$) from WNM-dominated ($\eta > 1$) regimes. The observed value in the 3C196 field is $\eta_\mathrm{obs} = 0.72$. }
    \label{fig:eta}
\end{figure*}

Figure~\ref{fig:eta} displays $\eta$ as a function of $f_{\mathrm{CNM}}$ and $u_{\mathrm{rms}}$. A horizontal dashed line at $\eta = 1$ indicates the threshold separating WNM-dominated ($\eta > 1$) from CNM-dominated ($\eta < 1$) regimes. The value observed in the 3C196 field, $\eta_\mathrm{obs} = 0.72$, is also shown.
In the perpendicular configuration, $\eta$ reaches high values, particularly for simulations with low CNM fractions, exceeding $\eta=7$ in some cases, which indicates a strong WNM–Faraday alignment. As $f_{\mathrm{CNM}}$ increases, $\eta$ decreases but generally remains above 1, confirming the persistent dominance of the WNM. In contrast, the parallel configuration shows $\eta$ values consistently close to or below 1 across the entire range of CNM fractions, with a flat trend. Only at high CNM fractions ($f_{\mathrm{CNM}} \gtrsim 35\%$) does the correlation approach the observational regime where CNM dominates.

To further disentangle the contributions from each phase, the two bottom panels show the respective alignment of WNM and CNM structures with the Faraday map. These panels confirm that the WNM largely dominates the correlation in the perpendicular case, while the CNM only begins to compete in the parallel configuration at high CNM fractions. This suggests that while the presence of CNM is a necessary ingredient for achieving a CNM-dominated regime, it is not the primary driver of the correlation, which depends more critically on the relative orientation of the magnetic field.

In the configuration where the LOS is aligned with the mean magnetic field, the maximum values of $\Tilde{V}^{98\%}$ for the CNM and WNM phases follow remarkably similar trends. This suggests that the morphological structures of CNM and WNM are comparable in this case and may both trace the same Faraday structures. In contrast, the perpendicular configuration reveals a clear divergence between the CNM and WNM $\eta$ values, indicating that their structures differ significantly. As also supported by Fig.~\ref{fig: NmomentsY}, this behavior suggests that the WNM structures are more closely aligned with the Faraday morphology, whereas the CNM structures deviate more substantially. This asymmetry highlights the dominant role of WNM in shaping the observed Faraday features, particularly when the magnetic field lies in the plane of the sky.

\section{Discussion}
\label{sec: discussion}

\subsection{Observation features well reproduced}

The numerical experiment presented here, representing the multi-phase \ion{H}{I} post-processed using a classical prescription to estimate the local ionization fraction, is able to reproduce several features seen in observations of the diffuse ISM at high Galactic latitudes.
\begin{itemize}
\item The velocity dispersion ($\sigma_{1pc}=0.7-2.3\,$km$\,$s$^{-1}$) and CNM fractions  ($f_{\textrm{CNM}} = 17-46$\%) are representative of what is measured at 21\,cm in the diffuse ISM of the Solar neighborhood \citep{Marchal2021,Marchal2024}. 
\item The range of $F(\phi)$ values is comparable to the ones measured with LOFAR in the 3C196 field with average Faraday depth of $-2 < M_1 < 3\,$rad$\,$m$^{-2}$ and spread of the Faraday spectra of $M_1 \sim 1-2.5\,$rad$\,$m$^{-2}$. 
\item The simulations of our experiment produce a range of peak polarization intensity between $0$ and $12\,$K (Figures~\ref{fig: Pmax} and \ref{fig: rotated column densities}). The highest values might not be representative as they are obtained when $\vec{B}$ is exactly perpendicular to the LOS, but our experiment is indicating that such relatively low \ion{H}{I} column density ($\langle N_{\textrm{HI}} \rangle = 1.55 \times 10^{20}\,$cm$^{-2}$) can produce $P_{\textrm{max}}$ values of a few K easily, similar to what is observed in the 3C196 field \citep[$P_\textrm{max} \leq 7\,$K][]{Jelic2015}.
\item The synthetic Faraday cubes produced, including the effect of a proxy of the LoTSS $uv$-minimum, show elongated structures with no polarization, reminiscent of the observed depolarization canals \citep[e.g.][]{Erceg2022}.
\item The HOG analysis used to evaluate the correlation between \ion{H}{I} and Faraday structures is showing significant correlation of the CNM features with $F(\phi)$, like it has been revealed in the 3C196 field \citep{Bracco2020}. 
\end{itemize}

Our results are indicative that the physics implemented in such numerical experiment is capturing fundamental aspects of the physics of the diffuse ISM. Even though these experiments were designed to reproduce the diffuse neutral gas (\ion{H}{I}), this study indicates that a simple prescription for computing the ionization fraction (and therefore $n_e$) is reproducing several properties of the LOFAR data, including a correlation between CNM structures and the morphology of $F(\phi)$ maps. We recall that before the results of \cite{Vaneck2017, Jelic2018,Bracco2020} the $RM$ values and Faraday structures observed in low-frequency radio data were thought to originate mostly from thermal electrons located in the Warm Ionized Medium (WIM). Our numerical study indicates that even with its low ionization fraction ($\sim 10^{-2}$ in the WNM and $\sim 10^{-4}$ in the CNM ) the \ion{H}{I} gas is contributing significantly to the Faraday rotation signal, to levels strong enough to capture a correlation between $F(\phi)$ and the CNM 21\,cm emission. This is compatible with the results of \cite{boulanger2024} who, using a totally different approach based on UV absorption data, also concluded that electrons in the WNM are a major contributor to the Faraday depth. 

\subsection{CNM-Faraday tomographic data correlation}

Even though the numerical experiment presented here reveals a significant correlation between CNM and Faraday structures, when looked closely, we could not find a single example in our simulations set that would reproduce perfectly the relative correlation of CNM and WNM with the Faraday tomographic data (i.e., $\eta$).

The Faraday depth values observed in the 3C196 field are almost centered on $0$ (see $M_1$ in Fig.~\ref{fig: histogram_M1 and M_2}-left). As mentioned in \cite{Jelic2018} this implies that the magnetic field is likely to be close to perpendicular to the LOS in this area of the sky. 
Indeed we reproduce this feature ($\langle M_1 \rangle \sim 0$) with synthetic observation with $\vec{B}$ perpendicular to the LOS, but the distribution of $M_1$ and the range of Faraday spread ($M_2\sim 1-2.5\,$rad$\,$m$^{-2}$) for this configuration are narrower than what is observed, even after de-biasing for noise.

Another important limitation concerns the physical size of the simulation box. The comparison between the simulated and observed Faraday depth distributions implicitly assumes that all polarized emission and rotation detected by LOFAR originate within a volume comparable to that of the simulation (50 pc on a side). In reality, the observed emission in the 3C196 field likely arises from a more extended region along the LOS, possibly spanning several hundred parsecs through the local magneto-ionic medium. This difference in physical depth may partly account for the narrower range of simulated Faraday depths and could also influence other derived quantities such as rotation measure dispersion and the \ion{H}{I}–polarization correlations. Nevertheless, achieving realistic CNM structures requires sub-parsec resolution, which cannot be maintained in simulations spanning Galactic-scale path lengths.

Regarding the \ion{H}{I}-Faraday correlation, the $\vec{B} \perp \textrm{LOS}$ case is the one for which the values of $\eta$ departs the most from the observations. As shown in Figures~\ref{fig: HOG_7 perp mean field} and \ref{fig: eta ratio} the CNM shows significantly smaller values of $\Tilde{V}$ than the WNM (i.e., higher values of $\eta$) than in the case where $\vec{B}$ is parallel to the LOS. The only way to obtain $\eta$ values close to the ones of the 3C196 field is by adding unrealistically large noise in the 21\,cm synthetic data, 10 times the one present in the real EBHIS data, in order to affect significantly the 21\,cm WNM signal but not as much the CNM. 
The case with $\vec{B} \parallel \textrm{LOS}$ produces values of $\eta$ that are closer to the observed ones, but this configuration shows values of $M_1$ that are significantly larger than the observations. 

CNM-$F(\phi)$ correlation (i.e. lower values of $\eta$) could increase by: An increase of $n_e$ in the dense \ion{H}{I} would produce a stronger Faraday signature associated to CNM structures. This could be adjusted by modifying the post-processing prescription used to estimate $n_e$ locally. 
Another way to increase the CNM correlation using HOG would be a stronger morphological correlation between the CNM filamentary structures and the orientation of $\vec{B}$. We have noticed in our simulations that many CNM clouds seem to have a random orientation with respect to $\vec{B}$ (see Fig.~\ref{fig: LIC along and perp}). This might require another forcing scheme for the simulations, maybe one using stellar feedback (c.f. supernovae) that might introduce coherent high pressure fronts over larger scales. 

Finally another way to lower $\eta$ would be to have a lower correlation between the WNM and Faraday structure. This could be caused by the WIM itself that we do not model in our setup. The WIM is likely to have a diffuse structure, similar to the WNM, with more large scale fluctuations. If the WIM is not spatially correlated to the WNM, its contribution to the Faraday depth is likely to affect more the WNM-Faraday correlation than the CNM one. The same is true if one consider the fact that LOS in the 3C196 field is likely to be about 4 times larger than what is modeled here. Adding up WNM signal along the LOS could lower the morphological correlation between WNM and Faraday. As the CNM occupies smaller volumes along the LOS, it is less affected by the projection effect of long lines of sight. All these effects are beyond the scope of this paper but they could be tested with dedicated numerical setups. Finally, our comparison relies entirely on a single LOFAR field and thus it samples a limited volume of the local interstellar medium. The magnetic and thermodynamic conditions along this line of sight may not be representative of other regions of the sky. Consequently, any correspondence between simulations and observations should be interpreted as illustrative rather than universally representative of the Galactic ISM that could be tested with the whole LoTSS mosaic \citep[e.g.,][]{Erceg2022, Erceg2024}.

\subsection{Depolarization canals}

A second aspect that is not well reproduced by our numerical setup are depolarization canals.
We observe them in our simulations and some seem to be caused by beam depolarization. On the other hand the morphology of the synthetic depolarization canals do not match the observations. Depolarization canals observed in the 3C196 field and over large areas at high Galactic latitude \citep{Erceg2022} are often very elongated, with axis ratios up to 10-100. The depolarization canals produced in our instrumental experiment are more twisty and relatively short.

This calls for another numerical study where the properties of the magnetic field should be varied more (we used a single value for the initial $\vec{B}$) as well as the properties of the mechanism injecting kinetic energy in the system. One might want to explore supernova feedback or alternatively Fourier driving with different timescales and/or with a more consistent orientation. In general turbulence-in-a-box simulations have a hard time to produce very elongated CNM features even in cases where they are found to be aligned with $\vec{B}$ \citep{Inoue&Inoutsuka16}.

\section{Conclusion} 
\label{sec: conclusions}

The main goal of this study was to evaluate if current state-of-the-art MHD numerical experiments used to model the thermally bistable \ion{H}{I} are able to reproduce properties of the Faraday sky and in particular of the \ion{H}{I}-Faraday correlation observed in the LOFAR data \citep[e.g.][]{Bracco2020}. 
Our parametric study is based on 50 pc simulations boxes (MHD + cooling) with a range of turbulent velocity forcing with different amplitude and level of compressibility. The forcing led to realistic ranges of velocity dispersion, CNM fraction and magnetic field fluctuations, reproducing values observed in the diffuse ISM. 
From this set we have produced synthetic 21 cm and synchrotron observations, including the effect of noise and the LoTSS $uv$-minimum, exploring a range of orientations of the line-of-sight from parallel to perpendicular to the mean $\vec{B}$ field. 

Our study reveals that thermal electrons associated with the \ion{H}{I} phase could contribute a significant fraction of the low frequencies (100-200\,MHz) synchrotron polarized emission observed at high Galactic latitudes. Indeed our synthetic Faraday observations reproduce levels of polarization intensity and $RM$ values that are commensurate with what is observed with LOFAR in the 3C196 field, our test region. 

To study in more details the contribution of CNM structures to the Faraday tomographic structures, we have developed a metric, $\eta$, based on the HOG algorithm \citep{Soler19}, that measures the ratio of WNM to CNM spatial correlation between 21 cm velocity channels and polarized intensity at specific Faraday depths. We have shown that this metric is quite robust with respect to the kernel size of the HOG algorithm.
Our analysis reveals a significant correlation of the CNM with Faraday structures, similar to what is observed in the 3C196 field \citep{Bracco2020}.
In details, the level of CNM-$F(\phi)$ correlation depends on several factors. 
\begin{itemize}
\item Simulations with low turbulent motions tend to have a stronger \ion{H}{I}-Faraday correlation in general, and also smaller $\eta$. 
\item The CNM-$F(\phi)$ correlation level depends on the $\vec{B}-\textrm{LOS}$ orientation; $\vec{B}$ perpendicular (parallel) to the LOS shows the lowest (highest) CNM-$F(\phi)$ correlation. 
\item Larger CNM-$F(\phi)$ correlation can be measured in faint and noisy 21\,cm data as higher 21\,cm noise tend to affect more the broad and low brightness WNM compared to the sharper and brighter CNM components. 
\item Noise in Faraday data reduces the level of correlation of all \ion{H}{I} phases equivalently, while not affecting $\eta$. 
\item Finally the level of CNM-$F(\phi)$ correlation does not depend strongly on $f_{\textrm{CNM}}$. We observe a very mild increase in CNM-$F(\phi)$ correlation with $f_{\textrm{CNM}}$, and only for $f_{\textrm{CNM}}<20\%$ and in the case where the LOS is perpendicular to the mean $\vec{B}$ field. 
\end{itemize}

Within the space we have explored, the CNM is always correlated with the Faraday tomographic structures but we could not reproduce exactly values of $\eta$ as low as found in the 3C196 field ($\eta=0.72$). This is indicative that CNM structures are more correlated to Faraday structures in the observations than what we could reproduced with our numerical experiment. This is specifically true if we consider simulations with physical conditions similar to the 3C196 field ($\vec{B} \perp \textrm{LOS}$, $\sigma_{u\textrm{,1pc}}=2.0\,$km\,s$^{-1}$ and $f_{\textrm{CNM}} = 38$\%). In these conditions we find $\eta \sim 1.5$ which indicates that WNM contributes more to Faraday structures than the CNM. 
Several elements could contribute to this result: the choice of the $n_e$ prescription, the contribution of the WIM to the Faraday depths (not modeled), the longer length of the LOS in the data compared to the simulations, the Fourier driving that tends to produce CNM structures not aligned with $\vec{B}$. 

Our results are indicative that low frequencies observations could serve as a probe of the magnetic field intensity and morphology in the neutral edges of the Local Bubble. Moreover, the detailed comparison of simulations and observations in the 21\,cm-Faraday space revealed subtle discrepancies that call for an improvement of the way the multi-phase, turbulent, magnetized and partly ionized ISM is being modeled today.

Finally, we note that our conclusions are drawn from the comparison with a single observational field. Future work should extend this analysis to additional LOFAR fields to test the robustness and generality of the trends identified here across different Galactic environments.

\bibliographystyle{aa}
\bibliography{refs}

\begin{appendix}

\section{MOOSE}
\label{app: MOOSE}
Mock Observation of Synchrotron Emission (\texttt{MOOSE}) is an interactive Julia-based tool designed to process numerical simulations of synchrotron emission and Faraday rotation. It generates synthetic observations by integrating three-dimensional cubes of physical parameters, including density, temperature, and magnetic field components. The electron density is computed using either empirical prescriptions or direct scaling with the neutral hydrogen density.

The code reconstructs polarization properties by computing synchrotron emission and applying Faraday rotation effects along the line of sight. It accounts for observational constraints by filtering large-scale structures to match interferometric data, convolving signals with telescope beam responses, and incorporating noise models to replicate real observations.

\texttt{MOOSE} employs 2D interpolation methods to efficiently estimate synchrotron emissivities based on precomputed lookup tables. The interpolation grid is currently designed to use LOFAR frequencies but can be adapted to any frequency range depending on the observational requirements. This technique allows for accurate integration of synchrotron radiation over a range of magnetic field strengths and frequencies while maintaining computational efficiency.

The code first reconstructs the Stokes parameters ($I_\nu$, $Q_\nu$, $U_\nu$) as a function of frequency before applying \texttt{RM-Synthesis} to derive the Faraday spectrum. This approach enables the decomposition of polarized emission along different Faraday depths, making it possible to analyze the magneto-ionic properties of the ISM.

The tool assumes an optically thin medium, meaning absorption effects are neglected. While this assumption is generally valid for high Galactic latitude regions, it may introduce biases in denser environments.

The synthetic data can be analyzed to extract rotation measure maps, Faraday spectra, and polarization structures.

\begin{figure}[!ht] 
\centering 
\includegraphics[width=\linewidth]{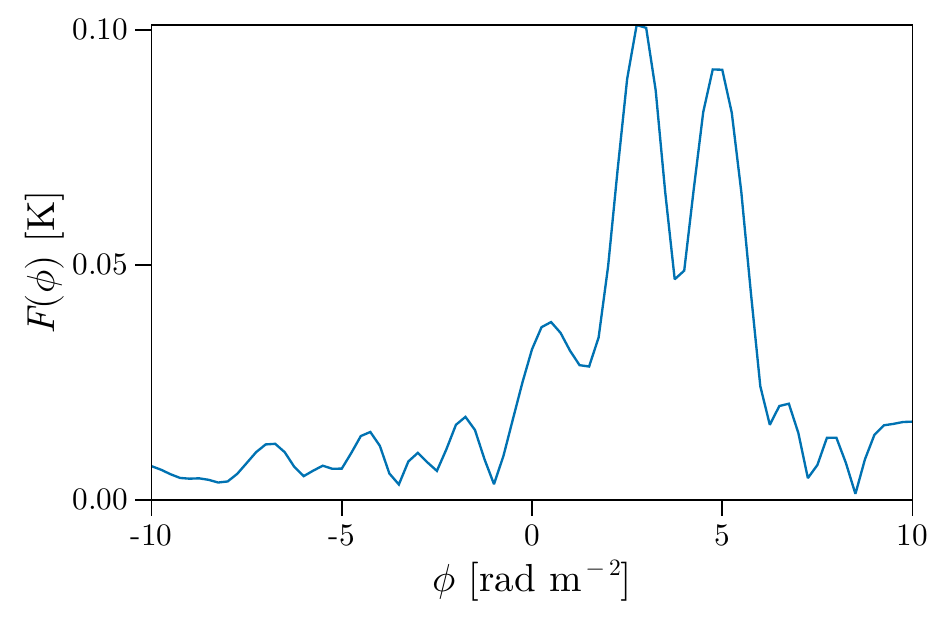}
\caption{Example of a Faraday spectrum produced by \texttt{MOOSE}. The x-axis represents the Faraday depth in $\textrm{rad m}^{-2}$, while the y-axis corresponds to the polarized intensity. The two primary peaks are located at $\phi_1 = 2.75~\textrm{rad m}^{-2}$ and $\phi_2 = 5~\textrm{rad m}^{-2}$.}
\label{fig: faraday spectrum}  
\end{figure}

Figure~\ref{fig: faraday spectrum} represents the polarized intensity as a function of the Faraday depth (in $\textrm{rad m}^2$). Peaks in the spectrum indicate distinct polarized emission components along the line of sight. The two most significant peaks in the spectrum correspond to Faraday depths of approximately $\phi_1 = 2.75~\textrm{rad m}^{-2}$ and $\phi_2 = 5~\textrm{rad m}^{-2}$, which likely indicate separate emitting regions along the line of sight.

It provides a visual representation of how \texttt{MOOSE} reconstructs the Faraday spectrum from input simulation data. The presence of sidelobes in the spectrum is a common feature resulting from the limited frequency coverage and resolution in \texttt{RM-Synthesis}. These sidelobes can introduce ambiguities in identifying true polarized components.

\section{Faraday moments}
\label{sec: Moments and effective width}
$M_0$ is the polarized intensity integrated over the Faraday depth range. $M_1$ is the polarized intensity weighted mean of Faraday depth $\phi$. $M_2$ is the width of the Faraday spectrum. These are analogous to moments in velocity space and are fully described in \citet{Dickey2019}. These quantities are affected by noise; moments are often estimated by thresholding the data, for instance \cite{Erceg2022} used only values six times higher than the noise level. Here we propose alternatives to estimate the Faraday moments without rejecting any data.  

Because $F(\phi)$ is a strictly positive quantity, noise in $F(\phi)$ has a specific effect. Noise fluctuations do not cancel out in calculating $M_0$ but they add up introducing a bias. For a given noise level $\sigma_{\textrm{noise}}$ (computed here by taking the standard deviation at the first 30 and the last 30 Faraday depths $\phi$) the noise bias on the first moment is equal to $N\,\sigma_{\textrm{noise}}$ where $N$ is the number of Faraday depths. To take this into account we compute the first moment as
\begin{equation}
\label{eq:M0}
M_0 = \sum_{i=1}^N (F(\phi_i) - \sigma_{\textrm{noise}})\, d\phi
\end{equation}
where $d\phi$ is the channel width in $\phi$.

The second-order moment $M_2$ is also affected by noise but it is also very sensitive to the presence of sidelobes in Faraday spectra (see an example in Fig.~\ref{fig: faraday spectrum}. These sidelobes arise from the limited range of $\lambda^2$ over which the Fourier transform is computed (see Eq.~\ref{eq: Fphi}). 
The spectral shape of $F(\phi)$ also depends significantly on the spectral shape of $Q_\nu$ and $U_\nu$ that can vary significantly due to Faraday rotation (Equations~\ref{eq: Unu} and \ref{eq: Qnu}). This results in spectral structures in $P(\lambda^2)$ that can vary significantly from one position to the next. These spectral features in $P(\lambda^2)$ are not instrumental as they depend on the physical distribution of emitting and rotating structures along the LOS \citep{Takahashi2023}. In certain conditions the combination of $Q_\nu$ and $U_\nu$ can even cancels signal in $P(\lambda^2)$.
As a consequence, the width of the Faraday spectrum can, in some cases, be smaller than the width of the RMSF, making its interpretation difficult.

As an alternative, we estimated the width of the Faraday spectrum using the effective width defined as
\begin{equation}
\label{eq:Weff}
W_\mathrm{eff} = \frac{M_0}{\max\limits_\phi F(\phi)}.
\end{equation}
This quantity reduces the influence of sidelobes by emphasizing the integrated power of the spectrum relative to its peak intensity. We found that $W_\mathrm{eff}$ provides a more robust and interpretable characterization of the spectral width in the presence of complex interference patterns. It is also not biased by noise like a standard second moment is. We computed $W_\textrm{eff}$ using $M_0$ corrected for the noise bias. Then the second order moment is estimated as
\begin{equation}
M_2 = \frac{W_\mathrm{eff}}{\sqrt{2\pi}}
\end{equation}
which is the exact solution for a Gaussian function.

Finally, the first moment $M_1$ defined as 
\begin{equation}
M_1 = \frac{\sum_{i=1}^N \phi_i \, F(\phi_i)}{M_0}
\end{equation}
is also biased in the presence of noise. 
As noise increases, the numerator tends to produce $M_1 \rightarrow \langle \phi_i \rangle$. If the $\phi$ vector is centered on $0$, like it is the case in our study, lower signal-to-noise ratio $F(\phi)$ will tend to have $M_1 \rightarrow 0$.
To circumvent this effect, we are using a matched filter approach. We leverage on the fact that $W_\textrm{eff}$ is a non biased estimate of the width of the signal, especially if the complexity of the Faraday spectra is small (i.e., $F(\phi)$ is characterized by one dominating component, large or narrow). The first moment is found by convolving $F(\phi)$ by a Gaussian filter with a width given by $W_\textrm{eff}$. $M_1$ is then estimated at the value of $\phi$ where this convolution is maximum.

\section{HOG results for the whole set of simulations}
\label{app: HOG14}
\begin{figure*}
\centering
\includegraphics[width=\linewidth]{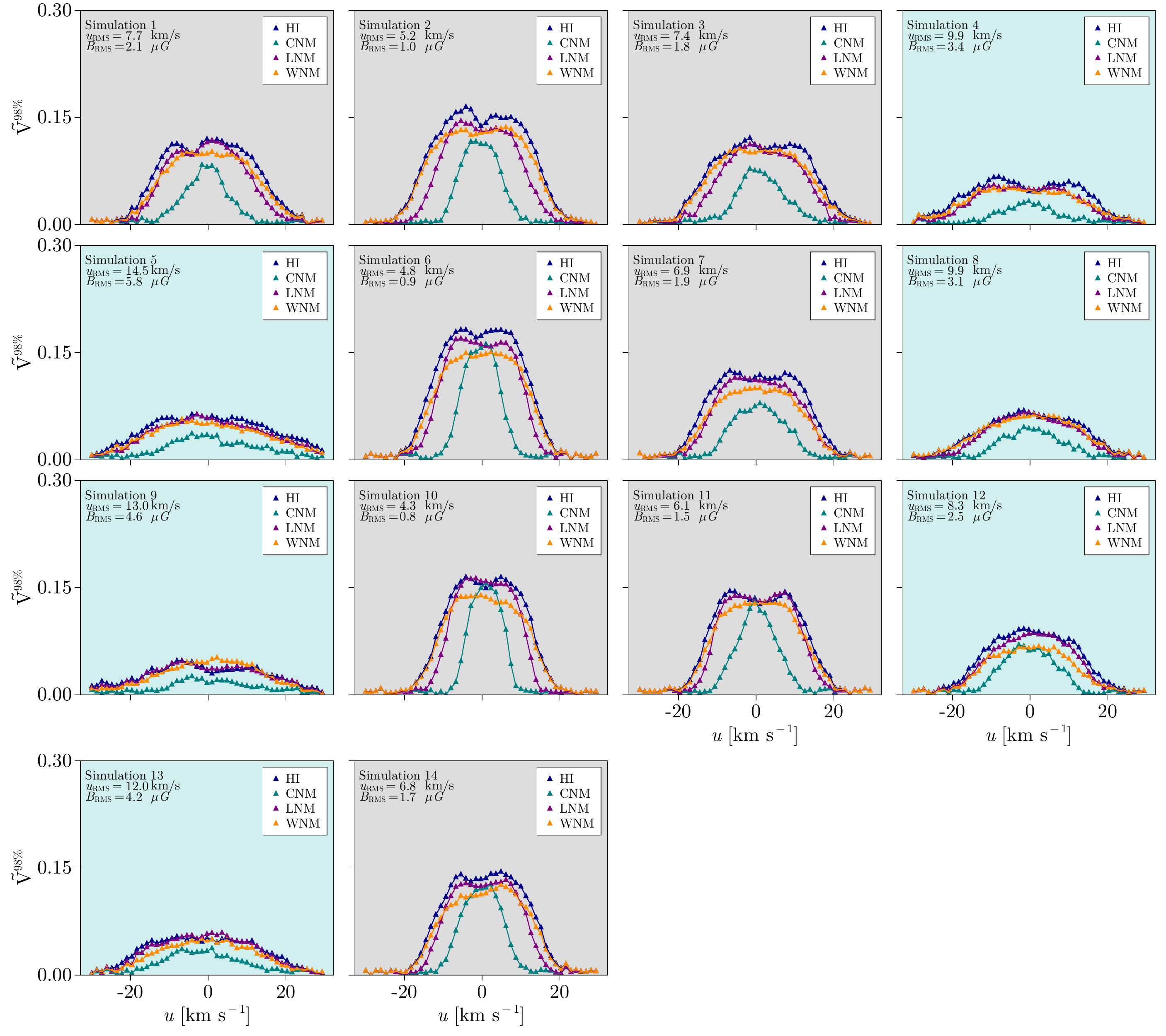}
    \caption{HOG results for all the set of simulations for the orientation of the LOS parallel to mean $\vec{B}$. In each panel the curve $\tilde{V}^{98\%}$ vs $u$ is shown for all \ion{H}{I} phases. Light blue backgrounds are simulations with higher turbulence ($v_\textrm{RMS}>8~\textrm{km s}^{-1}$) corresponding to higher magnetic field dispersion ($B_\textrm{RMS}>2.5~\mu\textrm{G}$). Gray backgrounds are the complement.}
    \label{fig: paraVoverVmax_allsimu_para}
\end{figure*}

\begin{figure*}
\centering
\includegraphics[width=\linewidth]{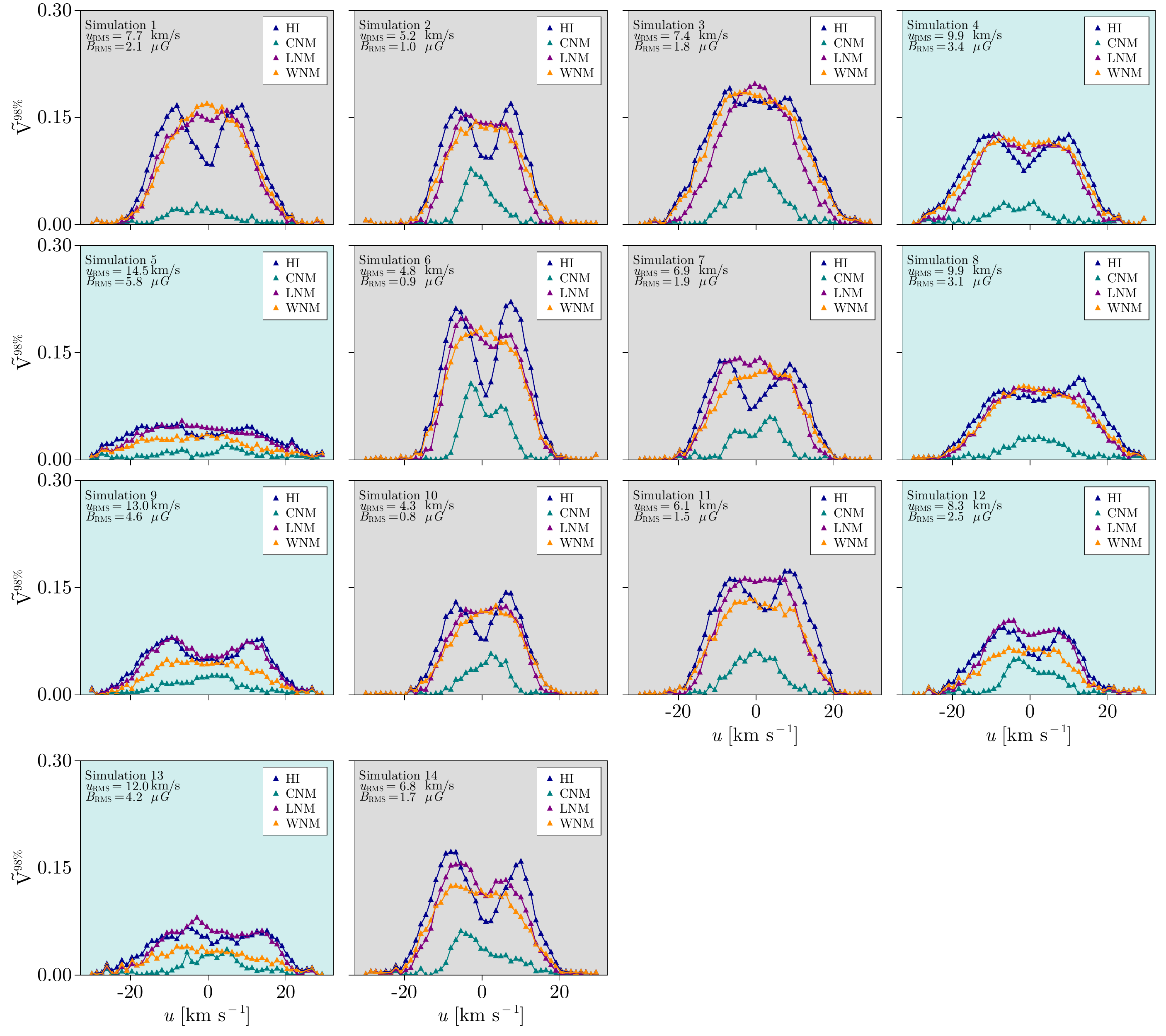}
    \caption{Same as Fig.~\ref{fig: paraVoverVmax_allsimu_para} but for cases where the LOS is perpendicular to mean $\vec{B}$.}
    \label{fig: paraVoverVmax_allsimu_perp}
\end{figure*}

The results of the HOG analysis for the whole 14 simulations set are presented in Fig.~\ref{fig: paraVoverVmax_allsimu_para} ($B$ parallel to the LOS) and Fig.~\ref{fig: paraVoverVmax_allsimu_perp} ($B$ perpendicular to the LOS). Different color backgrounds are used to illustrate low  ($v_\textrm{RMS}<8~\textrm{km s}^{-1}$ - gray background) and high ($v_\textrm{RMS}>8~\textrm{km s}^{-1}$ - light blue background) turbulence level. 

\section{Rotation of the cubes}
\label{app: Rotation}
To analyze physical quantities along different lines of sight, it is necessary to rotate the simulation cubes containing both scalar and vector fields. The rotation is particularly important in the context of these simulations because the magnetic field has a preferred orientation with a strong mean component in a specific direction. In our simulations, the ratio of the mean field strength to the root-mean-square fluctuation ($B/ B_{\text{rms}}$) is large, meaning that the field is strongly ordered. As a result, analyzing only the original simulation orientation provides a limited perspective on the system.

By rotating the cubes to intermediate angles, it becomes possible to explore configurations that are not perfectly aligned with the dominant field direction. This is crucial for understanding how turbulence and ordered structures interact under different projection effects. It also allows for a more comprehensive analysis of how observables change with different viewing angles.

\subsection*{Rotation of scalar fields}

For scalar fields such as electron density ($n_e$), temperature ($T$), and neutral hydrogen density ($n_H$), the rotation consists of a coordinate transformation from the original frame $(x, y, z)$ to the new frame $(x', y', z')$. Given a rotation matrix $R$, the new coordinates are obtained as:

\begin{equation}
\begin{bmatrix} x' \\ y' \\ z' \end{bmatrix} = R \begin{bmatrix} x \\ y \\ z \end{bmatrix}
\end{equation}

The scalar field values remain unchanged at each transformed coordinate. In practice, the rotation is implemented by interpolating the values of the scalar field onto the rotated grid using bilinear interpolation.

\subsection*{Rotation of vector fields}
For vector fields such as the magnetic field components ($B_x, B_y, B_z$) and velocity components ($V_x, V_y, V_z$), both the coordinate system and the vector components must be rotated. If the original vector field is represented as:

\begin{equation}
\vec{U} = \begin{bmatrix} U_x \\ U_y \\ U_z \end{bmatrix},
\end{equation}

then after rotation, the new components are given by:

\begin{equation}
  \vec{U'} = R \vec{U},  
\end{equation}

where $R$ is the same $3 \times 3$ orthogonal rotation matrix applied to the coordinate system. However, to correctly rotate a vector field, it is necessary to first apply the inverse transformation to determine where in the original coordinate system the vector should be sampled, and then apply the rotation to the vector itself:

\begin{equation}
    \vec{U'}(\vec{r'}) = R \vec{U}(R^{-1} \vec{r'}).
\end{equation}

These transformations ensure that the rotated vector components remain physically meaningful and aligned with the new coordinate system. By considering different rotation angles, we can analyze intermediate cases between the dominant field-aligned configuration and a purely turbulent regime. 

\begin{figure*}[!ht] 
\centering 
\includegraphics[width=\linewidth]{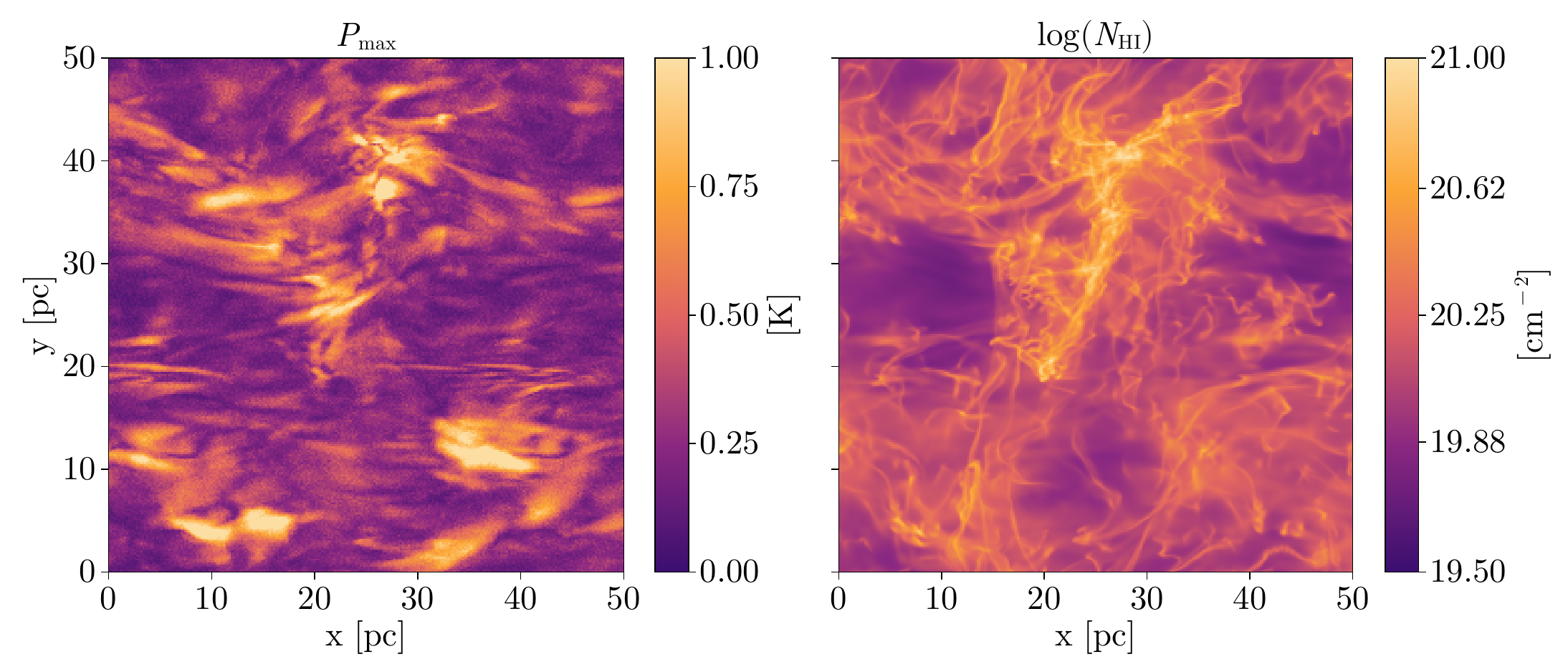}
\caption{\textit{Left:} Maximum polarized intensity of the rotated configuration. \textit{Right:} Column density of the same rotated configuration.} 
\label{fig: rotated column densities} 
\end{figure*}

\begin{figure*}[!ht] 
\centering 
\includegraphics[width=\linewidth]{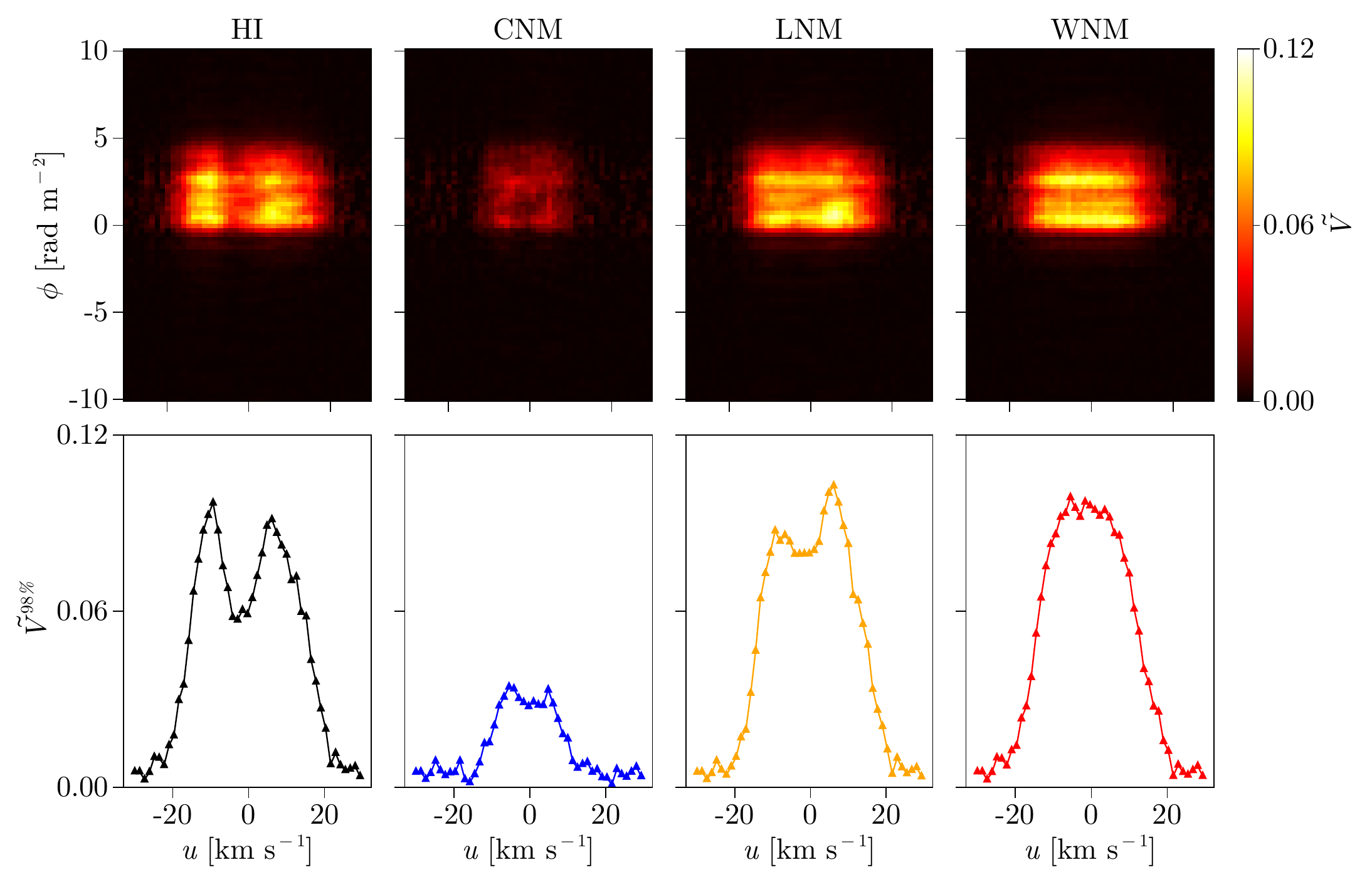}
\caption{HOG results for the rotated configuration of the case-study simulation similar to Fig.~\ref{fig: HOG_7 along mean field}. CNM is still less correlated with Faraday structures than the WNM. Nevertheless, concerning the values of $\Tilde{V}$, these are similar to the LoTSS data.} 
\label{fig: HOG_rotated} 
\end{figure*}

Figure~\ref{fig: rotated column densities} shows how looks like the rotated $P_\textrm{max}$ and column density when we rotate the simulation cubes around the $z$-axis by a $45^\circ$. Fig.~\ref{fig: HOG_rotated} the result of HOG after the same rotation.

\end{appendix}
\label{LastPage}
\end{document}